\documentclass[aps,prb,twocolumn,superscriptaddress]{revtex4}
\usepackage{epsfig}
\usepackage{color}
\usepackage{hyperref}
\usepackage{graphicx}

\usepackage{amsmath}
\usepackage{comment}

\emergencystretch=\maxdimen
\begin{document}
 
\title{Stabilization of singlet hole-doped state in infinite-layer nickelate superconductors} 

\author{Mi Jiang} 
\affiliation{Institute of Theoretical and Applied Physics, Jiangsu Key Laboratory of Thin Films, School of Physical Science and Technology, Soochow University, Suzhou 215006, China} 

\author{Mona Berciu}
\affiliation{Department of Physics and Astronomy, University of
British Columbia, Vancouver B.C. V6T 1Z1, Canada}
\affiliation{Stewart Blusson Quantum Matter Institute, University of
  British Columbia, Vancouver B.C. V6T 1Z4, Canada}

\author{George A. Sawatzky}
\affiliation{Department of Physics and Astronomy, University of
British Columbia, Vancouver B.C. V6T 1Z1, Canada}
\affiliation{Stewart Blusson Quantum Matter Institute, University of
British Columbia, Vancouver B.C. V6T 1Z4, Canada}

\begin{abstract} 
Motivated by the recent X-ray absorption spectroscopy (XAS) and resonant inelastic X-ray scattering (RIXS) experiments, we use a detailed impurity model to explore the nature of the parent compound and hole doped states of (La, Nd, Pr)NiO$_2$ by including the crystal field splitting, the Ni-$3d$ multiplet structure, and the hybridization between Ni-$3d$, O-$2p$, and Nd-$5d$ orbitals. For simplicity and stimulated by the recent electronic structure calculations, the latter are formally replaced with symmetric orbitals centered at the missing O sites in the Nd layer, forming a two-dimensional (2D) band strongly hybridizing with the Ni-$3d^9_{z^2}$ state. 
This hybridization pushes the main part of the $3d^9_{z^2}$ spectral function up in energy by several eV and stabilizes the singlet with considerable $d^9_{z^2}$ and other configurational components. 
For the parent compound, we find that states of Ni-$3d^9_{z^2}$ character spread over a large energy range in the spectra, and cannot and should not be represented by a single orbital energy, as suggested in other approximations. This is qualitatively consistent with the RIXS measurements showing a broad distribution of the Ni-$3d^9_{z^2}$ hole state, although the shape of the Ni-$3d^9_{z^2}$ related structure is much more complicated requiring reinterpretations of the RIXS data.  For the hole-doped systems, we show that adding these additional ingredients can still result in the lowest-energy hole doped state having a singlet character.  
\end{abstract}

\maketitle

\section{Introduction} 
The recent discovery of superconductivity (SC) below a critical temperature $T_c\sim 15$K in  Nd$_{0.8}$Sr$_{0.2}$NiO$_2$ thin films~\cite{2019Nature} initiated studies of the new family of Ni-based superconductors.\cite{Aritareview,Botana_review,Pr,LaSC,Nd6Ni5O12} One promising strategy is to use of ``reasoning-by-analogy'' to achieve better understanding of unconventional superconductivity in other SC families, especially the high-T$_c$ cuprates. 
Because the unusual Ni$^{1+}$ oxidation state has the same $3d^9$ electronic configuration like Cu$^{2+}$, the infinite NiO$_2$ planes were naively expected to host similar properties with the CuO$_2$ planes.  The initial excitement about drawing this parallel between cuprate and nickelate superconductors has cooled off, however, owing to various theoretical and experimental findings which indicate that the newly found nickelate superconductors show important difference from the cuprates~\cite{Aritareview,Botana_review, Held2022,Hanghuireview,Hwang2022}.

From an experimental point of view,  the bottleneck in this rapidly evolving research field lies in the difficulty of synthesis and characterization of these new nickelates with the unusual Ni$^{+}$ ions~\cite{2019Nature,Hall1,Hall2,sample2,Hwang2022}. Even so, there are already many findings that are proving difficult to combine in a coherent picture. The upturn of the resistivity at low temperatures~\cite{2019Nature,Pr,Hwang2022} suggests a (still debated) possible involvement of Kondo physics,~\cite{Hwang2022,GuangMing} while the normal state can be treated either as a bad metal or a weak insulator~\cite{Botana_review,Hwang2022}. The parent compound appears to not host long-range magnetic order~\cite{Pickett2004,Hepting,dft1,dft9,dft16,dft18,dft26,dft25,noLRO} in spite of having magnetic correlations~\cite{mag1,mag2,mag3,mag4,mag5}. Hall coefficient measurements show that, at low temperatures, the charge carriers switch from electrons in the parent compounds to holes in the superconducting and over-doped systems.\cite{Pr,Hall1,Hall2,Hwang2022} This is taken as  evidence of the multi-orbital character of the infinite-layer nickelates~\cite{Hepting}, although there are also proposals supporting the single-band picture.\cite{oneband,Held2022} The nature of the superconducting pairing is undoubtedly the  feature of most interest in the literature. A recent single particle tunneling study revealed the spatial coexistence of $d$-wave and $s$-wave~\cite{dswave}, while a recent London penetration measurement strongly challenged the $d$-wave pairing scenario by supporting a predominantly nodeless pairing~\cite{nodeless2022}. 
We note that several theoretical studies support the scenario of spin fluctuations as the glue for  $d$-wave superconductivity, similar to  cuprates,~\cite{dft2,dft3,dft101,dft102,spinfluc1,spinfluc2} although our earlier work~\cite{Mi2020} suggested that the superexchange interaction  in nickelates is decreased by about one order of magnitude, compared to cuprates.

Many of the theoretical investigations attempting to understand the differences between cuprates and the infinite-layer nickelates use density functional theory (DFT),~\cite{Pickett2004,dft1,dft2,dft3,dft4,dft5,dft6,dft7,dft8,dft9,dft10,dft24,dft26} also in combination with dynamical mean-field theory,  DFT+DMFT,~\cite{dft101,dft102,dft11,dft12,dft13,dft14,dft15,dft16,dft17,dft18,dft19,dft20,dft21,dft22,dft23,dft25,Held2022,Hanghuireview}  to calculate the electronic structure so as to uncover the contributions from different orbitals to the important states near the Fermi energy. Most of these studies agree that one significant difference between the two classes of materials is the appearance in the nickelates of a rather broad band that crosses the Fermi energy and is composed of a combination of orbitals including Nd-$5d_{xy}$, Nd-$6s$, O-$2p$, Ni-$3d_{z^2}$, Ni-$4s$, and interstitial states. This can be seen clearly in projections of the density of states on the atomic orbitals upon which there is a lot of density missing in the interstitial region instead. 
This broad band is believed to be essential to explain fascinating properties such as the self-doping effect in the nickelate parent compound~\cite{GuangMing}, suppression of the magnetic order~\cite{Pickett2004,Hepting,dft1,dft9,dft16,dft18,dft26,dft25} etc. Its existence suggests very different low-energy physics in the two classes of materials.

In terms of identifying a reliable model Hamiltonian, the debate continues on whether the interplay between correlations and hybridizations  favors the Hubbard or the Hund mechanisms.~\cite{dft12,dft20,dft23} In a previous study~\cite{Mi2020} we argued that the Ni$^{1+}$O$_2$ layers fall inside a ``critical'' region and should be classified as  Mott-Hubbard insulators according to the ZSA classification~\cite{ZSA1985}, with a singlet hole-doped state of similar symmetry with that in CuO$_2$; this has been supported by a few recent experiments~\cite{Goodge2020,oneband}. In this view, the doped hole is primarily located in a linear combination with $x^2-y^2$ symmetry of neighbor ligand O orbitals. This is very different from having the doped hole primarily occupying the $3d_{z^2}$ orbital, as in a Hund's rule favored triplet state. Clearly, establishing which of these very different scenarios is relevant will have significant bearing on the debate about similarities (or lack thereof) between the two classes of superconductors.

Recent X-ray absorption spectroscopy (XAS) and resonant inelastic X-ray scattering (RIXS) experiments~\cite{oneband} are interpreted to show that the doped holes dominantly reside in the $d_{x^2-y^2}$ orbital, partially supporting the single band Hubbard model scenario~\cite{dft101,dft4,dft24,Held2022}. However, they also show a  significant $z$-polarized character indicating the presence of Ni-$3d_{z^2}$ holes in the undoped ground state. This is not consistent with our previous findings. However, in that work we ignored the existence of the broad band crossing the Fermi energy, which is known to have strong hybridization with the Ni-$3d_{z^2}$ orbitals. Therefore, it is important to understand how its inclusion affects our results.  

This is why here we  build on our previous work by adding more ingredients to our model, to understand their relevance. We again start from a multi-orbital model of infinite-layer nickelates~\cite{Mi2020} studied with an impurity approximation, including the local Ni-$3d$ multiplet structure of all $d$ orbitals, to investigate its corresponding undoped and hole doped ground states. In particular, we focus on how the ``critical'' character of the doped hole singlet state found previously, can be affected  in  more realistic settings. Specifically, we investigate  (i) the effect of including crystal field splittings of the $3d$ levels, and (ii)  the role played by hybridization between the NiO$_2$ plane and the electronic states in its two neighbour planes of Nd atoms giving rise to the broad band crossing $E_F$. As already mentioned, this band arises from a complicated mix of many orbitals. 
Here, we avoid this complexity by using instead single $s$ symmetry orbitals Zs centered at the O vacancy positions in the Nd layer, consistent with the recently proposed electride-like behaviour of the infinite-layer nickelates~\cite{Kat}. 

We find that inclusion of {\em only} crystal field splittings can push the $3d_{z^2}$ level to the higher energies observed in experiments, but it remains a very narrow peak. Inclusion of the hybridization between this orbital and the broad band is necessary in order to see it spreading over a wide energy range. With both new ingredients added in the model, we continue to find an undoped ground-state with primarily $3d^9_{x^2-y^2}$ character, and a hole-doped state with $^1A_1$ singlet character similar to the cuprates. But now we also find there is a strong $3d_{z^2}$ involved in this lowest energy singlet state, which also strongly involves the s vacancy band state. The presence of $3d_{z^2}$ holes in the lowest energy hole-doped state makes Ni look more like Ni$^{2+}$ but apparently low spin, which is consistent with RIXS and XAS results although a more careful analysis of the experimental results is needed first. 

The paper is organized as follows: Section II discusses the multi-orbital Ni impurity model and the formalism used to find its spectrum. Section III illustrates various results of spectral functions, phase diagram, ground state composition etc. both in the absence and in the presence of the hybridization between Ni-3d and the effective ``zeronium'' Zs band in the Nd layer. Finally,  Section IV summarizes our findings and provides further perspective.

\section{Model and methodology}

\begin{figure}[t!]
\psfig{figure=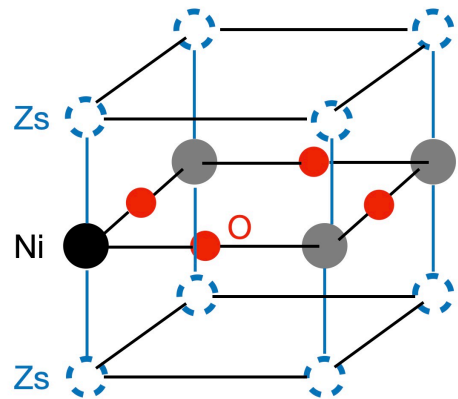,width=0.8\columnwidth,clip=true, trim = 0.0cm 5.0cm 0.0cm 5.0cm} \\
\caption{Schematic view of the atoms involved in our model impurity calculation. The NiO$_2$ layer is treated within an impurity approximation of one Ni (black sphere) embedded in the O square lattice (red circles; only four O are depicted but we include the full O lattice). The other Ni ions (gray circles) are ignored. The adjacent layers of Nd are modelled by hybridization between Ni orbitals and the ``zeronium'' states labeled Zs (dashed circles). See text for more details. }
\label{geom}
\end{figure}

Before introducing our Hamiltonian, it is useful to explain its underlying assumptions.  We begin from a single Ni$^{1+}$-$3d^9$ impurity embedded in an infinite square lattice of O-$2p^6$ ions, {\it i.e.} the problem studied in Ref.~\onlinecite{Mi2020}. First, we supplement that work by including crystal field splittings of the $3d$ orbitals to understand their effect both on the resulting undoped ground-state, and also on the one-hole doped state ({\em i.e} when we remove one more hole from the system described above, for a total of two holes missing from otherwise filled orbitals). Note that this splitting is a result of the ionic charges on the Nd ions producing a substantial crystal field, as also reported in the quantum chemistry calculations~\cite{Alavi}. The ligand field splitting due to the  orbital dependent hybridizations with the O-2p and the Zs states are already part of our model Hamiltonian.  

More substantially, we then also include hybridization with the states in the Nd plane concentrating on those involved in the highly dispersive band crossing the Fermi energy, as seen in most DFT and DFT+U calculations. We model this by a single $s$ symmetry orbital centered at each O vacancy position Zs (depicted in Fig.~\ref{geom}) and follows the recent suggestion that the infinite-layer nicklelates have properties similar to those of electrides~\cite{Kat}. This approach is reasonable because the actual orbital character of that band has comparable components of the various Nd, O and especially Ni-$3d_{z^2}$ character, which is consistent with the recent work indicating that the dominant hybridization between Ni-$3d$ orbitals and itinerant electrons in the rare-earth spacer layer is through this interstitial $s$-like orbital, due to a large inter-cell hopping~\cite{dft25}. Similarly, our ab-initio calculations confirm that there is this so-called ``Zeronium'' band (spatially centered at these O vacancies) crossing the Fermi energy~\cite{Kat}. The band structure calculation shows that there are also orbitals with $p$ symmetry involved in the creation of this vacancy, however those only have weak $\pi$-bonding with the $d_{xz/yz}$ orbitals, which is why we ignore them. To summarize, the O vacancy is treated as an $s$-like orbital located appropriately in the Nd layer, which has  appreciable hybridization with the Ni-$3d_{z^2}$ orbital~\cite{dft25,Kat}. As we show below, this extra complication involving Ni-$3d_{z^2}$ orbital has a dramatic influence on the spectral functions of the various states.


The corresponding Hamiltonian, then, is:
\begin{equation}
\label{H}
{\cal H} = E_{s} + K_{pd} + K_{pp} + K_{ds} + K_{ss} + V_{dd}.
\end{equation}
Here,
\begin{equation}
\label{H1}
E_s = \sum_{m\sigma} \epsilon_d(m) d^\dagger_{m\sigma}d^{\phantom\dagger}_{m\sigma} 
	  + \sum_{in\sigma} \epsilon_s s^\dagger_{i\sigma}s^{\phantom\dagger}_{i\sigma}
      + \sum_{jn\sigma} \epsilon_p p^\dagger_{jn\sigma}p^{\phantom\dagger}_{jn\sigma} 
\end{equation}
describes the on-site energies of the various orbitals included in the calculation. Specifically, $d^\dagger_{m \sigma}$ creates a {\em hole} with spin $\sigma$ in the Ni-$3d_m$ orbital, with an associated energy $\epsilon_d(m)$;  $p^{\dagger}_{jn \sigma}$ creates a  {\em hole} with spin $\sigma$ in the orbital O-$2p_{n}$, $n\in \{x,y,z\}$,  located at site $j$ of the O sublattice, with a corresponding energy $\epsilon_p$; and $s^\dagger_{i \sigma}$ creates an {\em electron} with spin $\sigma$ at the Zs site $i$ in a neighbor Nd layer. In our previous work we set $\epsilon_d(m)=0$,\cite{Mi2020} but here we allow for finite crystal field splitting, motivated by the recent XAS/RIXS experiments~\cite{oneband}. As further explained below, we set $\epsilon_d({x^2-y^2})=0$ and adjust the remaining crystal field splittings until the one-hole  spectra (characterizing the undoped parent compound) are consistent with experimental findings. 

Before continuing, it is important to emphasize that we use a dual language, with hole excitations to describe the configuration of the nearly filled Ni and O orbitals, and electron excitations to describe the almost empty band of Zs states.  The former choice follows up on our previous work,\cite{Mi2019} while the latter choice is because in agreement with many other DFT studies,  our first principle calculations~\cite{Kat} revealed that the Zs band of the undoped parent NdNiO$_2$ is almost empty, with quite low electron occupancy of $\sim 0.03$/unit cell. This is why it is sensible to count the electrons in this nearly empty band. 

The hybridization between the various $m$ orbitals of the Ni impurity and the various $n$ orbitals of its 4 nearest neighbour (NN) O sites located at the sites $\langle .j\rangle$ is described by:
\begin{equation}
\label{H2}
K_{pd} = \sum_{\langle .j\rangle mn \sigma} 
(T^{pd}_{mn} d^\dagger_{m \sigma}p^{\phantom\dagger}_{jn \sigma}+h.c.)
\end{equation}
while the hopping between various nn O orbitals is given by:
\begin{equation}
\label{H3}
K_{pp} = \sum_{\langle jj'\rangle nn' \sigma} 
(T^{pp}_{nn'} p^\dagger_{jn \sigma}p^{\phantom\dagger}_{j'n' \sigma}+h.c.)
\end{equation}
The hopping integrals $T^{pd}_{mn}$ and $T^{pp}_{nn'}$ are determined following Slater and Koster~\cite{Mi2020,SlaterKoster}. In the following, we specify the values  for the {\em magnitudes} of the  $t_{pd\sigma}, t_{pd\pi}, t_{pp\sigma}, t_{pp\pi}$ hopping parameters, and note that the signs coming from the corresponding orbitals' overlaps are properly included in the Hamiltonians. 

The hybridization between the Ni $3d_m$ orbitals and its NN Zs orbitals located at the sites  $\langle .i\rangle$ is described by:
\begin{equation}
\label{H4}
K_{ds} = \sum_{\langle .i\rangle m \sigma} 
(T^{ds}_{m} d^\dagger_{m \sigma}s^{\dagger}_{i \sigma}+h.c.) 
\end{equation}

The first term describes the key new process where an electron hops from one of the Ni orbitals, thus creating a hole behind, into the (otherwise empty) Zs states. Strictly speaking, spin conservation imposes the combination $d^\dagger_{m \sigma}s^{\dagger}_{i, -\sigma}$, but we use the simpler notation because proper labelling of the electrons' spins is irrelevant in our impurity model. Again, below we give the  magnitude of this hopping as $t_{ds}$, and the proper signs are included in the Hamiltonian.

\begin{figure*}[t]
\psfig{figure=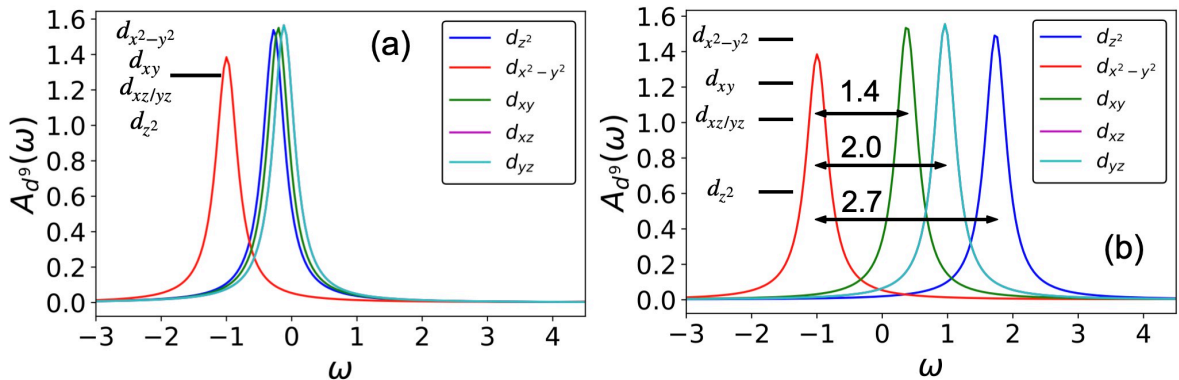, width=.9\textwidth,, trim = 0.0cm 6.0cm 0.0cm 6.0cm} 
\caption{Single hole spectra for the undoped parent compound in the {\em absence} of hybridization to Zs orbitals, for (a) $\epsilon_d(m)=0$ and (b) $\epsilon_d(m)$ tuned so as to obtain $d^9$ spectra consistent with XAS/RIXS experiments~\cite{oneband}. The inset of (b) shows the corresponding $\epsilon_d(m)$ in electron language. With respect to $\epsilon_d(x^2-y^2)=0$, the crystal fields are $  \epsilon_d(xy)=0.6$eV, $\epsilon_d(xz/yz)=1.1$eV, $\epsilon_d(z^2)=2.1$eV.}
\label{fig2}
\end{figure*}

Electron hopping between Zs orbitals is given by:
\begin{equation}
\label{H5}
K_{ss}= \sum_{\langle ii'\rangle \sigma} 
(T^{ss} s^\dagger_{i \sigma}s_{i' \sigma}+h.c.)
\end{equation}
and is characterized by a magnitude $t_{ss}$ for intra-layer hopping between NN Zs orbitals in each layer, and by   $t_{ss\perp}$ for inter-layer hopping between NN Zs orbitals located in the top and bottom layers.

Finally, 
\begin{equation} \label{H6}
V_{dd} = \sum_{\bar{m}_1\bar{m}_2\bar{m}_3\bar{m}_4} U(\bar{m}_1\bar{m}_2\bar{m}_3\bar{m}_4) d^\dagger_{\bar{m}_1}d^{\phantom\dagger}_{\bar{m}_2}d^\dagger_{\bar{m}_3}d^{\phantom\dagger}_{\bar{m}_4}
\end{equation}
describes correlations of the impurity Ni orbitals, with the shorthand notation $\bar{m}_x \equiv m_x \sigma_x$ where $x=1,\dots,5$ denotes spin-orbitals. To be more precise,  two-hole $3d^8$ configurations are naturally involved in our calculation, and require the consideration of the Coulomb and exchange interactions for all singlet/triplet irreducible representations of the $D_{4h}$ point group spanned by two $d$ holes, in terms of the Racah parameters $A$, $B$ and $C$ ~\cite{Mi2019,Mi2020}. In principle, inclusion of $d^7$ configurations with much more complicated interactions is also possible, but for simplicity we ignore them, because their energy would be at $\approx 3U$, where the Hubbard $U$ is estimated to be at least 8eV or higher (see below).

As discussed in our previous work~\cite{Mi2020}, the Ni-O and O-O hybridizations are estimated to be $t_{pd}\approx 1.3-1.5$ eV and $t_{pp}\approx0.55$ eV,~\cite{Kat,Si2020,dft15} on the same scale as in cuprates. 
Meanwhile, the Racah parameters $B, C$ are set by atomic physics so we keep the same values $B=0.15, C=0.58$ eV as in cuprates. One significant difference between  NiO$_2$ and CuO$_2$ are the charge transfer energies $\Delta(m)=\epsilon_p-\epsilon_d(m)$. For the $m=x^2-y^2$ orbital,  $\Delta$ is estimated to be $\Delta \approx 7-9$ eV in nickelates as opposed to $\Delta\approx 3$ eV in cuprates~\cite{Christensen}.

We are interested in the spectra corresponding to different configurations with various symmetries relevant to both the undoped (hosting one hole) and hole-doped (hosting two holes) infinite-layer nickelate. Without the inclusion of the Zs band, the configurations reduce to the single- and two-hole states discussed in our previous study~\cite{Mi2020,Mi2019}.  As discussed  in more detail below, we supplement these with  configurations that  describe ``self-doped'' states with an electron in the Zs band compensated by an additional hole in the Ni layer, due to the strong hybridization of the Ni-$3d_{z^2}$ orbital with the Zs states which results in spectral weight to $z$-polarized XAS spectra at the Ni-$2p$ edge. 

The spectra are extracted from the generalized propagators for each specific configuration. For example, $d^8$ spectra
$A^{\Gamma}(\omega)$ for a particular irreducible representation $\Gamma$ assuming that one hole has already occupied the Ni-$3d_{x^2-y^2}$ orbital reads 
\begin{align} \label{Aw}
& A^{\Gamma}(\omega) = -\frac{1}{\pi} \sum_{m} \lim_{\delta\rightarrow 0} \Im G_{d}(m,b_1,\omega+i\delta;\Gamma)
  \nonumber \\ & G_{d}(m,b_1,z;\Gamma) = \langle 0| d^{\phantom\dagger}_{b_1}
  d^{\phantom\dagger}_{m} \hat{G}(z) d^{\dagger}_{m} d^{\dagger}_{b_1} |0 \rangle
\end{align}

All the calculations of the propagators are performed by employing the variational exact diagonalization with standard Lanczos solver. 
The variational space is constructed by imposing a cutoff distance ${\bf R_c}$ between the holes/electrons. Obviously, $R_c \rightarrow \infty$ recovers the full Hilbert
space. We typically set $R_c>15$ for the results shown below. 
We also use a relatively large broadening $\delta$ to avoid the situation where continua in the spectra look like a collection of peaks. 

\section{Results}  

\begin{figure*}[t!]
\psfig{figure=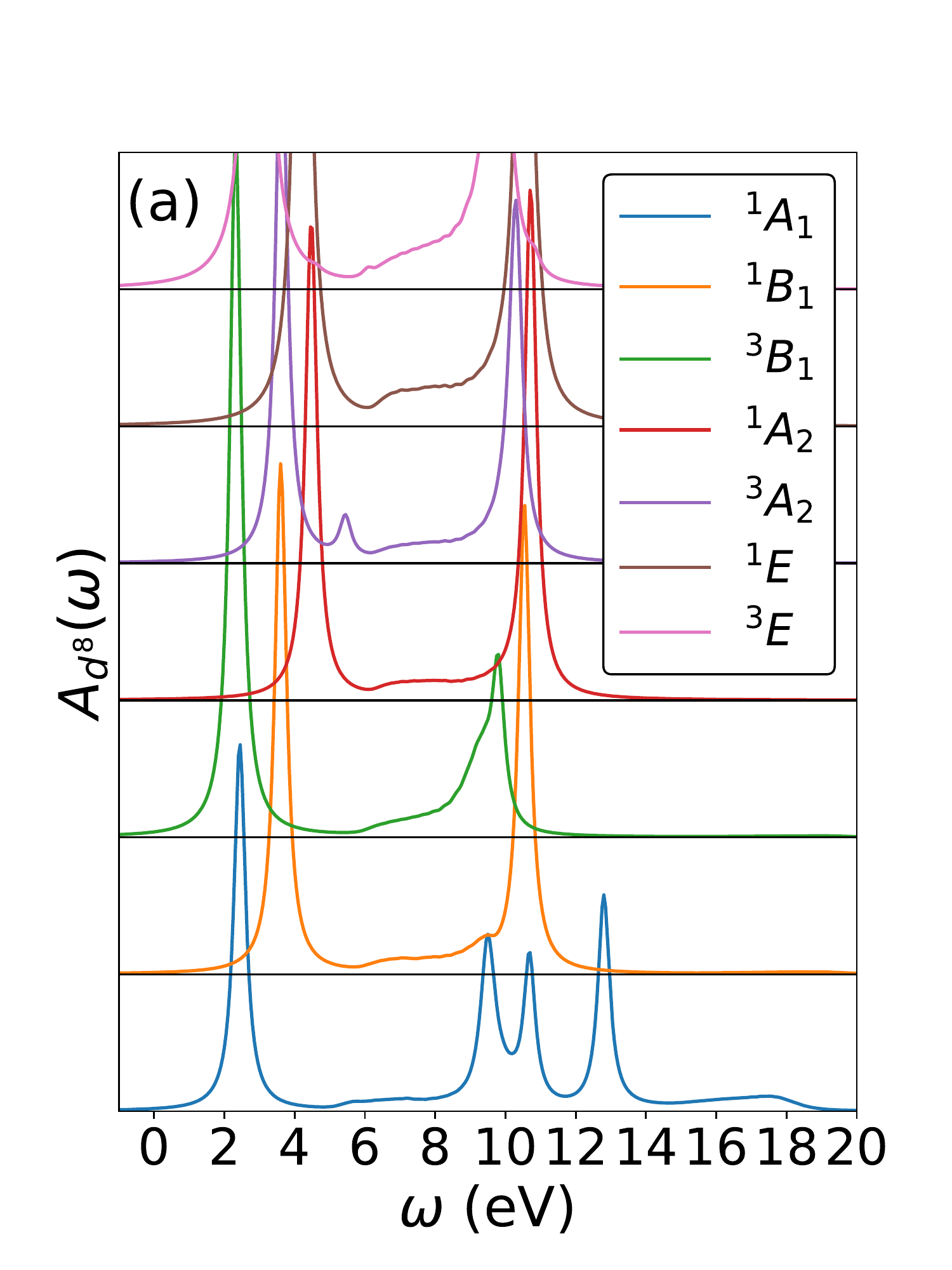, width=.42\textwidth,height=8.5cm,trim={0 0 0 2.4cm},angle=0,clip} 
\psfig{figure=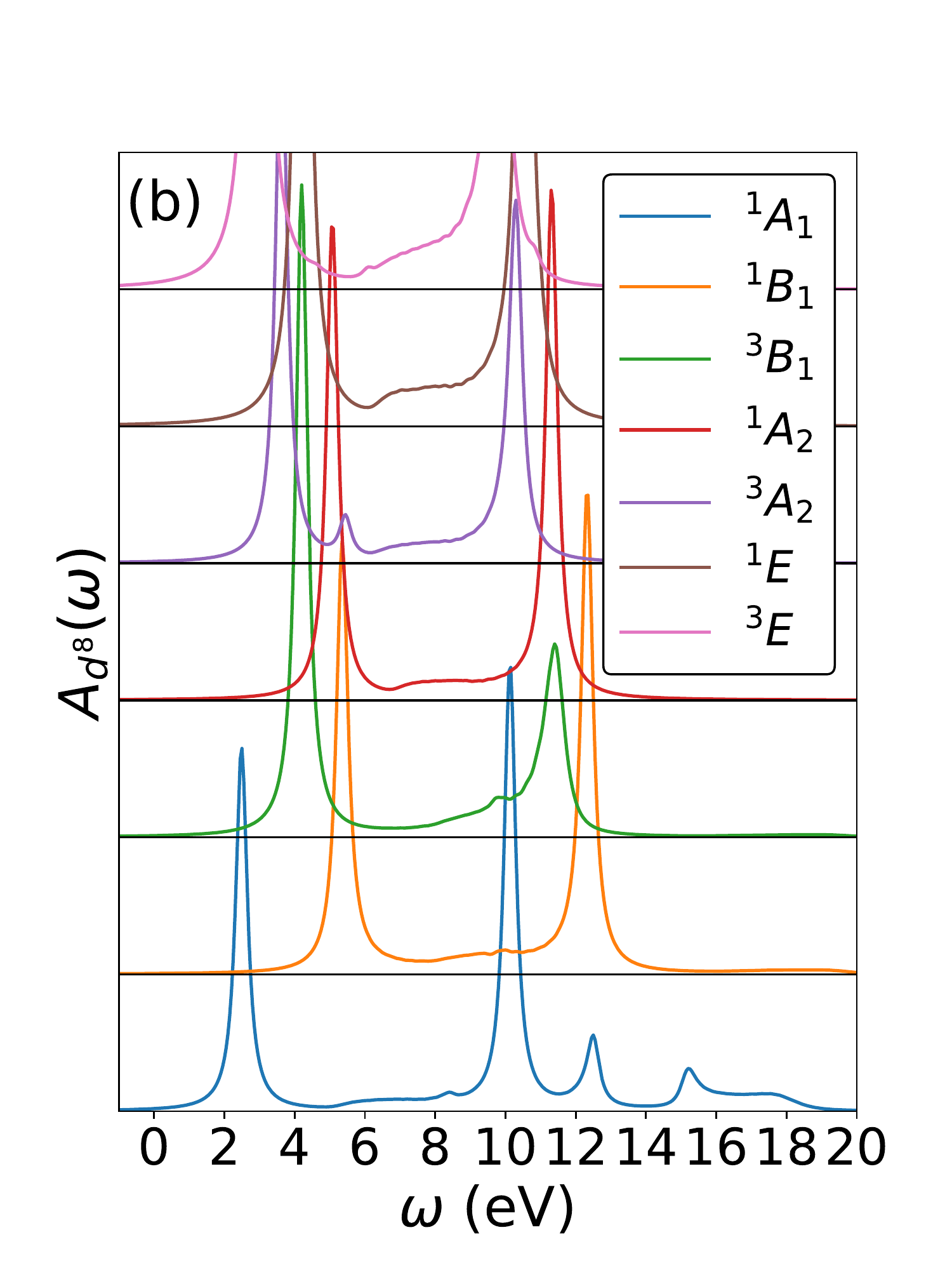, width=.42\textwidth,height=8.5cm,trim={0 0 0 2.4cm},angle=0,clip} \\
\psfig{figure=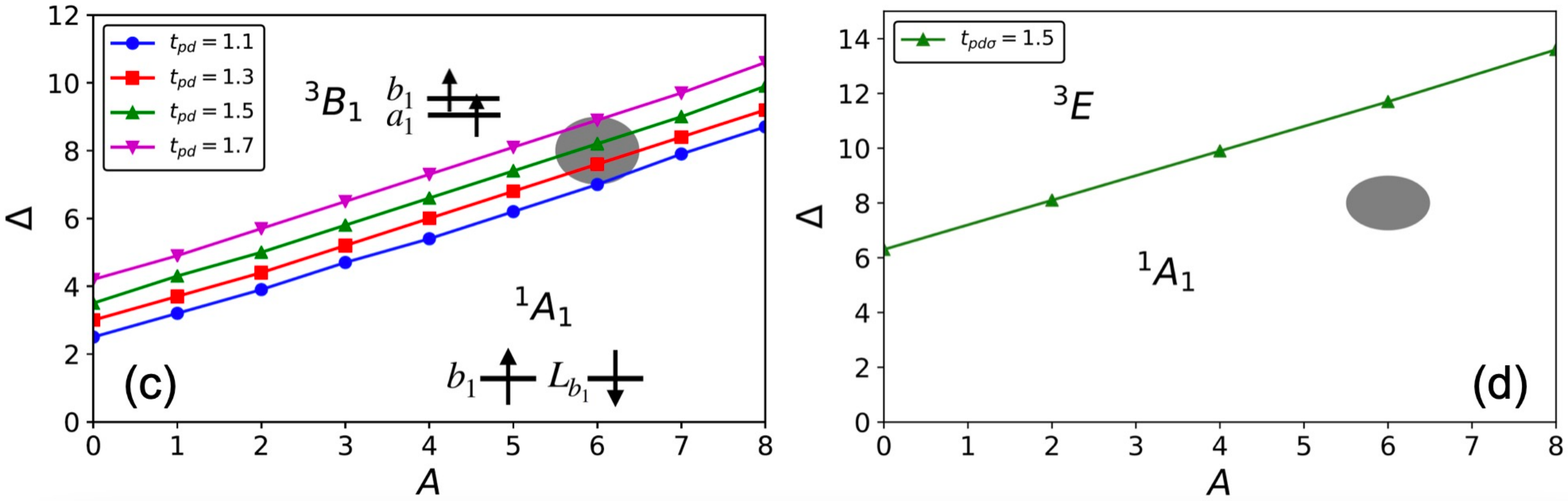, width=.9\textwidth,angle=0,, trim = 0.0cm 6.0cm 0.0cm 6.0cm} \\
\caption{(a-b) Two hole spectra (hole doped system) at $A=6.0$eV, $\Delta=8.5$eV in the {\em absence} of hybridization to Zs orbitals for (a) $\epsilon_d(m)=0$ and (b) tuned $\epsilon_d(m)$ (the same values as in Fig.~\ref{fig2}). The two-hole state with $^{1}A_1$ symmetry (ZRS-like) is stabilized by the additional crystal field splitting; (c-d) Two-hole (one hole doped) ground state phase diagram for $\epsilon_d(m)=0$ in (c), and for the tuned  $\epsilon_d(m)$ in (d). The shaded gray region is expected to be relevant for the infinite-layer nickelates. }
\label{fig2a}
\end{figure*}

Recent XAS/RIXS experiments found that the peaks corresponding to the $d_{xy}$, $d_{xz/yz}$ and $d_{z^2}$ orbitals in the parent compound are located at 1.4 eV, 2.0 eV and 2.7 eV from the $d_{x^2-y^2}$ peak, respectively.~\cite{oneband} These values disagree with what we obtained in the absence of crystal field splitting, {\em i.e.} when $\epsilon_d(m)=0$, in Ref. \onlinecite{Mi2020}. Those older results are reproduced in Fig. \ref{fig2}(a), which shows that in the absence of crystal fields, the $x^2-y^2$ peak is about 1eV below the other (nearly degenerate) peaks, due to its enhanced in-plane $pd$ hybridization.

\subsection{Tuning of the crystal field splittings}

Our first step is to find the values of the crystal field splittings $\epsilon_d(m)$ that allow us to produce peak locations in agreement with the XAS/RIXS data, in the {\em absence} of hybridization with the Nd layers. In this case, Eq.(1) is reduced to   the model 
 ${\cal H} = E_s+K_{pd}+K_{pp}+V_{dd}$ used in Ref. \onlinecite{Mi2020} plus a tunable crystal field splitting. 

A reasonable result is displayed in Fig.~\ref{fig2}(b). The inset shows the ordering of $\epsilon_d(m)$ in electron language: relative to $\epsilon_d(x^2-y^2)=0.0$, we find $\epsilon_d(z^2)=2.1, \epsilon_d(xy)=0.6, \epsilon_d(xz/yz)=1.1$eV. As expected, the peak with $d_{z^2}$ symmetry can be moved to the observed higher energy by sufficiently increasing its crystal field.

We then performed the two-hole calculations to obtain the $d^8$ spectra corresponding to the hole doped NiO$_2$  to find whether the hole-doped ground-state has triplet $^3B_1$ or singlet $^1A_1$ character.~\cite{Mi2020} Fig.~\ref{fig2a}(a-b) illustrate the change in the symmetry of the hole-doped ground state when tuning $\epsilon_d(m)$, for fixed $A=6.0$eV, $\Delta=8.5$eV.
In Fig.~\ref{fig2a}(c-d), we  draw the corresponding phase diagrams without and with the crystal fields. The presence of the crystal fields moves the transition line to higher $\Delta$ values, as shown by the comparison of the  $t_{pd\sigma}=1.5$ eV results. As a result, the gray area marking parameters relevant for nickelates, moves further inside the $^1A1$ region, making the infinite-layer nickelates more similar to the cuprates in terms of the nature of doped hole states. This is fully expected as well, because the $d_{z^2}$ hole is now much higher in energy. As a result, the Hund triplet consisting of two holes in $d_{z^2}$ and $d_{x^2-y^2}$ orbitals is even more energetically costly compared to the singlet state.

\subsection{Inclusion of Ni-Zs hybridization}
\begin{figure*}[t!]
\psfig{figure=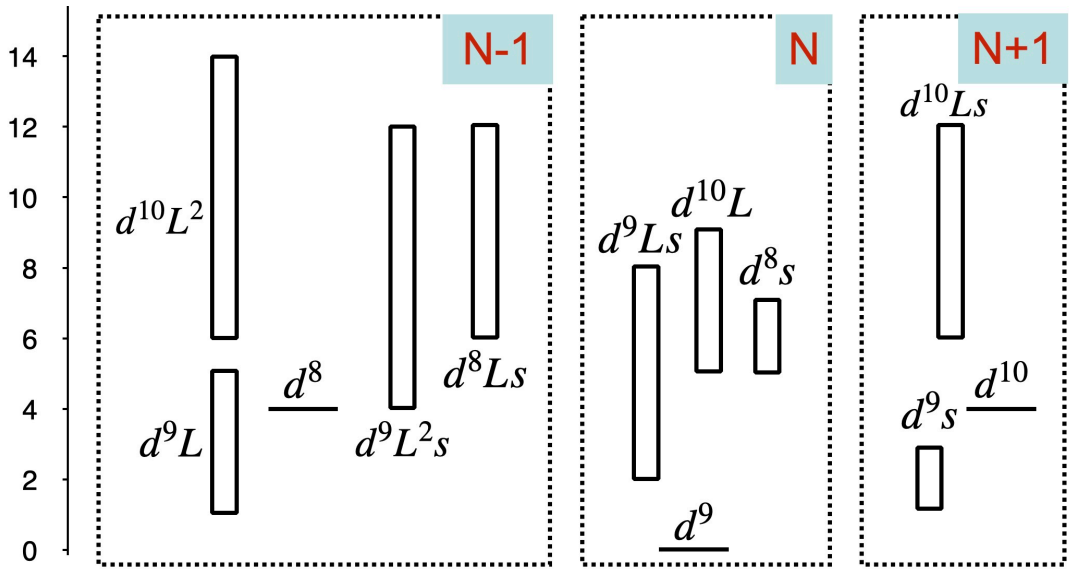,width=.8\textwidth, clip=true, trim = 0.0cm 3.0cm 0.0cm 0.0cm} 
\caption{The configuration energy level diagram (vertical scale is energy in eV estimated from DFT calculations) in the {\it absence of the Ni-O and Ni-Zs hybridizations}, for $U=8$eV, $\epsilon_d(m)=0$, $\epsilon_p=3$eV, $\epsilon_s=2$eV, $t_{pp}=0.5$eV, $t_{ss}=0.25$eV in Eq.~\ref{H}. The vacuum state is chosen to be Ni-$3d^{9}$ O-$2p^6$ Nd-$s^0$ in CS$_N$ (denoted as $d^9$ because of the absence of O's hole excitation and Zs's electron excitation). 
This choice is motivated by the convention of electronic structure calculation to split up the influence of the Hubbard $U$ on $d^8$ and $d^{10}$ by putting $U/2$ on the electron removal (CS$_{N-1}$) as well as $U/2$ on the electron addition (CS$_{N+1}$) states separately. Note that $d^8$ and $d^9$ represent d states of various symmetries and the zero energy is exact for the $d^9_{x^2-y^2}$ state.}
\label{fig3}
\end{figure*}

Although the single hole spectral peak positions can be tuned to match experiments, they are still sharp peaks. This is inconsistent with the experimental observation that the $d_{z^2}$ state spreads out over a large energy range. To obtain such a broad feature, it is necessary to couple the $d$ orbitals, especially $d_{z^2}$, to some other dispersive bands. The obvious choice are the bands associated with the Nd layers. Therefore, as described by the full Eq.~\ref{H}, from now on we replace the hybridization with Nd-$5d$ orbitals with that with an effective $s$ orbital centered at the O vacancy position Zs, which is also the position of the empty muffin-tin in the DFT calculations~\cite{dft25,Kat}.

Before showing the results, it is useful to review the complexity of this problem by identifying the various kinds of states spread over various energy ranges, that are mixed by hybridization to give rise to the relevant spectra. To avoid confusion, from now on we will use the configuration language to label these various states.

First, by using the procedure described in the original ZSA work~\cite{ZSA1985}, we define CS$_N$ to be the manifold including Ni-$3d^{9}$ O-$2p^6$ Zs-$s^0$ and all other configurations  (detailed below) connected to it via various hybridizations, at a fixed total number $N$ of electrons.  The lowest energy eigenstate obtained after diagonalizing the Hamiltonian of Eq. (\ref{H}) within CS$_N$ is the ground-state of the undoped infinite layer NdNiO$_2$ (within the single Ni-impurity approximation).  Similarly, CS$_{N-1}$ is the manifold for the hole-doped system, including all doublets Ni-$3d^{8}$ O-$2p^6$ Zs-$s^0$ and all other configurations connected to them through hybridization. The lowest eigenstate obtained after diagonalizing ${\cal H}$ within CS$_{N-1}$ will reveal the nature of the lowest energy state associated with a doped hole, in particular whether it is a singlet or triplet. For completeness, we also analyze the CS$_{N+1}$ manifold for the electron-doped system. It contains  Ni-$3d^{10}$ O-$2p^6$ Zs-$s^0$ and all other states connected to it through hybridizations.  
These states are important when comparing to the LDA+U calculations and are also needed for the interpretation of the XAS and RIXS data, which involve the $d^{10}$ state accompanied by a core hole. 

In Figure \ref{fig3} we sketch the various states in the CS$_N$ and CS$_{N\pm1}$ manifolds {\em in the absence of all Ni-O and Ni-Zs hybridizations}. For simplicity, here we ignore the crystal field effects and show all $d^9 \equiv$ Ni-$3d^{9}$ O-$2p^6$ Zs-$s^0$ states as having the same energy. Similarly, all multiplet splittings are ignored which is why all $d^8 \equiv$ Ni-$3d^{8}$ O-$2p^6$ Zs-$s^0$ configurations are shown as degenerate. 
The $d^9$ states of  CS$_N$ are chosen as the vacuum state of zero energy, while the $d^8$ and $d^{10}$ states are both placed at $U/2$. Recall that the Hubbard $U$ for the $3d$ levels is formally defined as $U=E(d^{10})+E(d^{8})-2E(d^{9})$. In terms of Racah parameters, $U = A+4B+3C \approx 8$ eV for our typical values.
We note that the results shown below do include both crystal fields and correlations of the Ni-$3d$ levels, so the degeneracies of the $d^9$ and $d^8$ states are lifted accordingly from their corresponding baselines sketched in Fig. \ref{fig3}.

Details of the states included in each manifold are as follows:

(i) {\bf The CS$_{N}$ manifold}: Starting from $d^9$, Ni-O hybridization allows an electron to hop from a neighbour O to the empty Ni-$3d$ orbital, resulting in $d^{10}L$ states ($L$ indicates a ligand hole in the O band). These were the only states included in our previous work.~\cite{Mi2020} They spread over the bandwidth $8 t_{pp}\approx 4$eV of the O band. $\Delta$ is measured from $L$ band's center at $\Delta=E(d^{10}L)-E(d^{9})=\epsilon_p+U/2$. Given the estimated $\Delta \sim 7$ eV and  typical value $U=8$ eV, we must therefore set $\epsilon_p=3$ eV.  

The Ni-Zs hybridization adds two other continua, by allowing hopping of an electron primarily from the Ni-$d_{z^2}$ orbital into the empty Zs band. The hopping integrals between other Ni-3d orbitals and Zs are all zero by symmetry, although the hopping between Ni-3d and the actual Nd-5d states are finite. This process generates the $d^8s$ states ($s$ denoting an electron in the zeronium s band) starting from the $d^9$ configuration. They are centered at $E(d^8s)-E(d^9)=U/2+ \epsilon_s$ and have a bandwidth $8t_{ss}$ of the zeronium band. We remark that $8t_{ss}$ would be the bandwidth for only hopping within a single Zs plane. In fact, we have also included the hopping between the two Zs planes sandwiching the NiO$_2$ layer so that the Zs bandwidth is further broadened in our realistic calculations. 
The sketch in Fig. \ref{fig3} sets $\epsilon_s=2$eV, but we will treat it as a free parameter in the following.
This hybridization also generates the continuum $d^9Ls$ starting from the $d^{10}L$ states. This is centered at $\epsilon_s+\epsilon_p$ and its bandwidth is the convolution of the O and Zs bands. 

We emphasize that the $d^9Ls$ states can only be reached from the $d^{10}L$ states through the hybridization between $d_{z^2}$ and Zs. This is important because it allows mixing with the $d^9$ states of both $x^2-y^2$ as well as of $3z^2-r^2$ character. The $d^8$ states that can be further reached must have at least one $z^2$ hole, which influences the energy of $d^8$ triplet states. We revisit these points below, where we analyze the results.

We ignore all higher energy states in this manifold such as $d^{10}L^2s, d^8Ls^2, d^7Ls$ etc., because their contribution to the ground-state is expected to be really small.

\begin{figure*}
  \psfig{figure=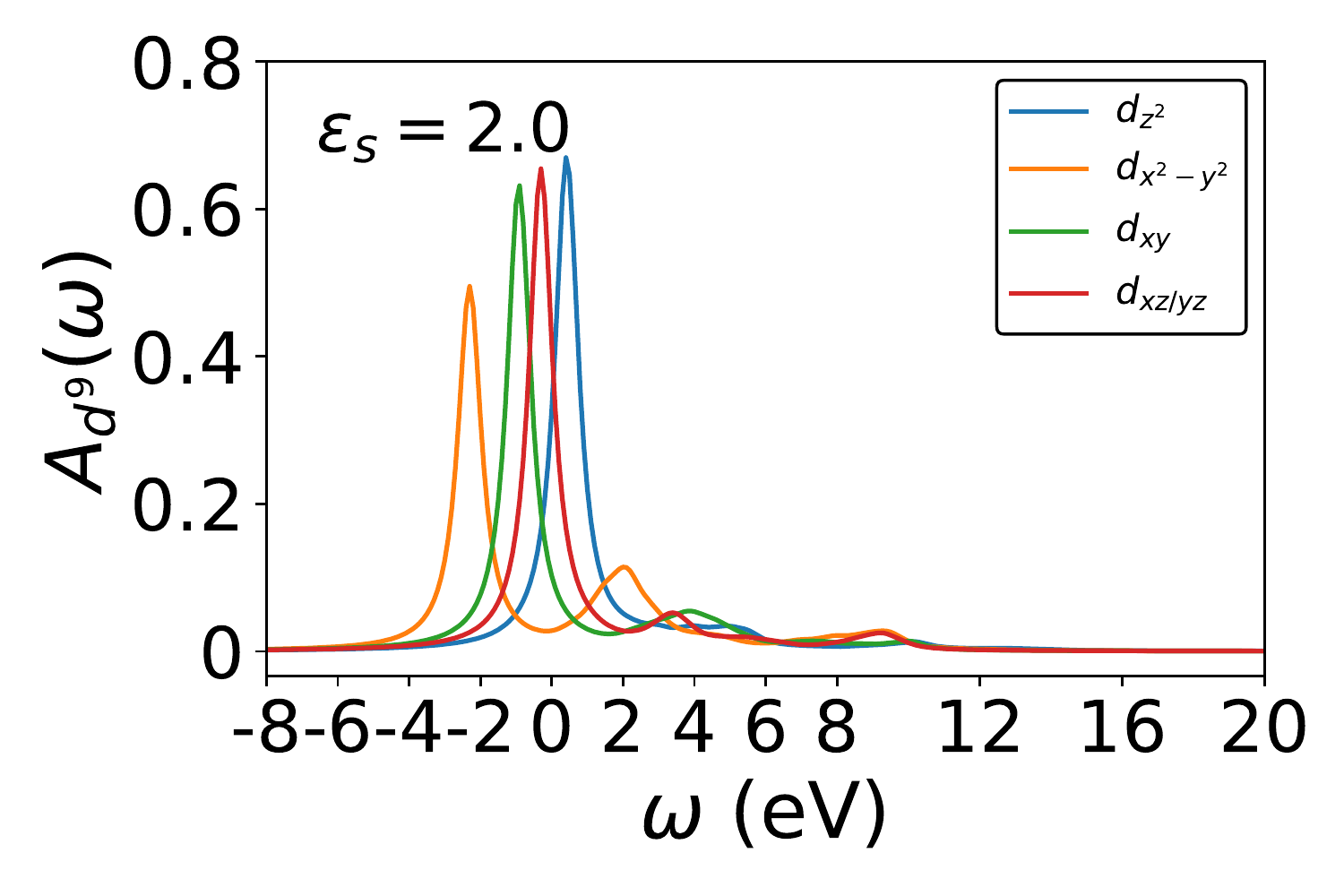,width=.65\columnwidth, trim={0 1.8cm 0 2.0cm}}
   \psfig{figure=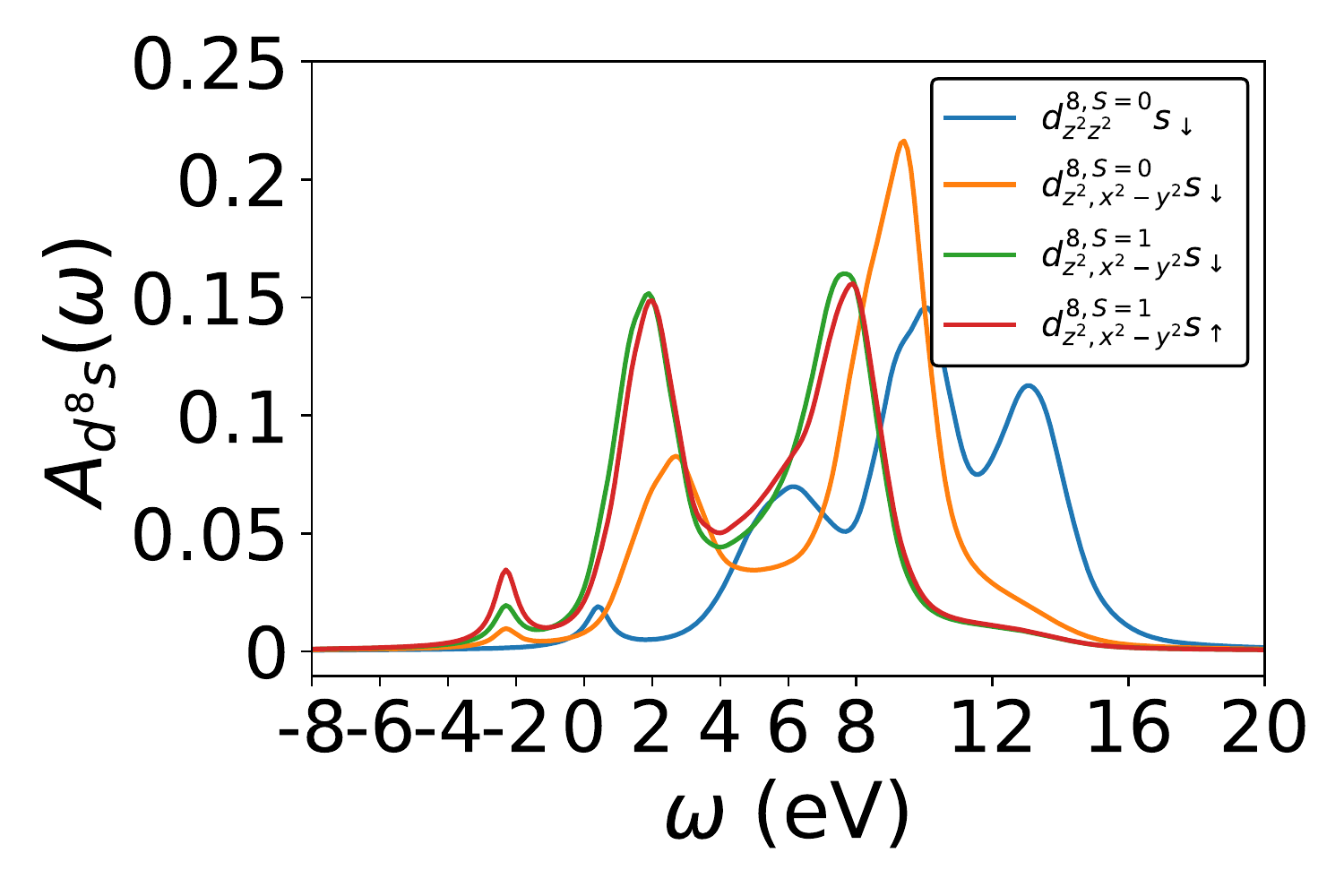,width=.65\columnwidth, trim={0 1.8cm 0 2.0cm}}
   \psfig{figure=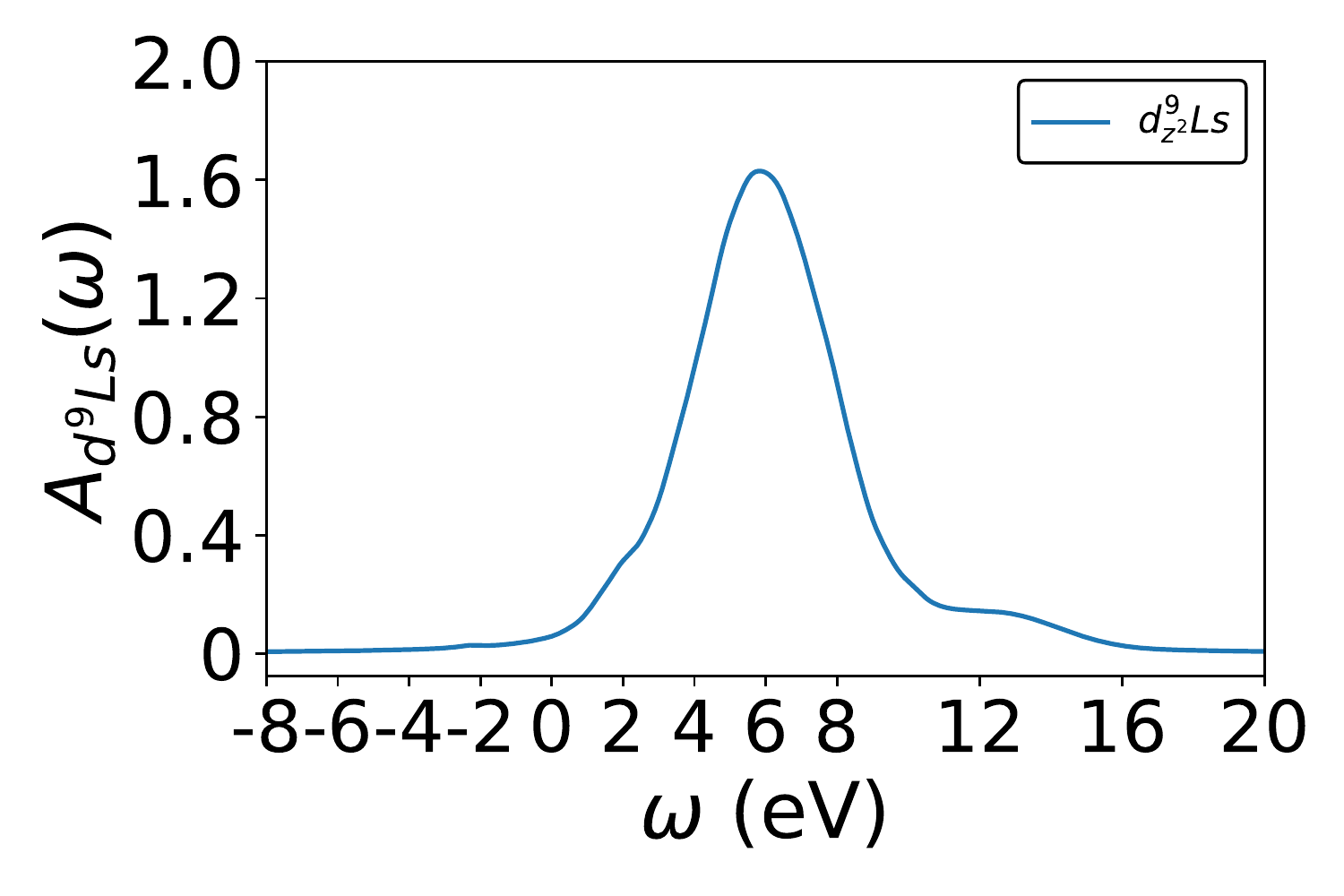,width=.65\columnwidth, trim={0 1.8cm 0 2.0cm}}
   \psfig{figure=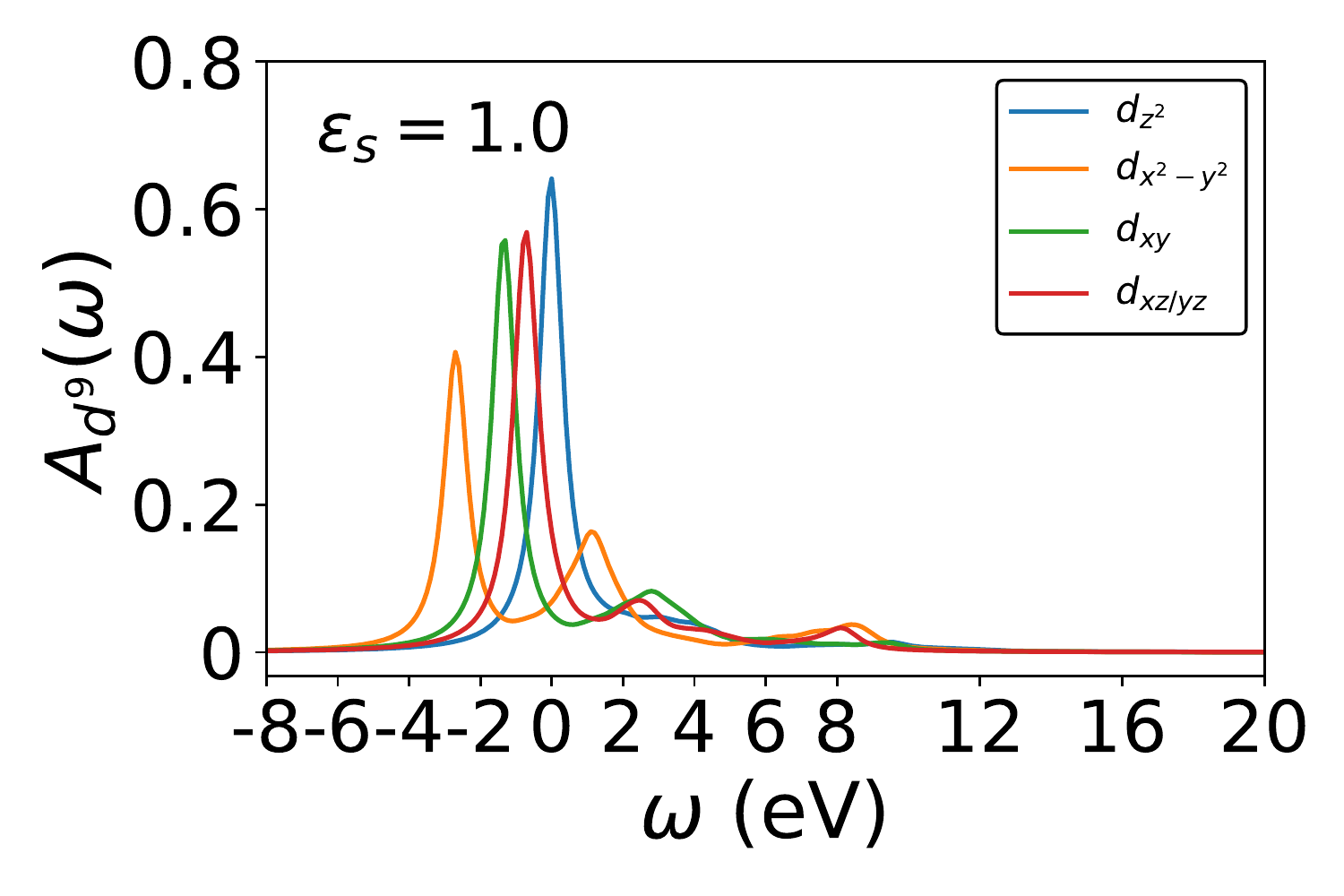,width=.65\columnwidth, trim={0 1.5cm 0 0.3cm}, clip=true}
  \psfig{figure=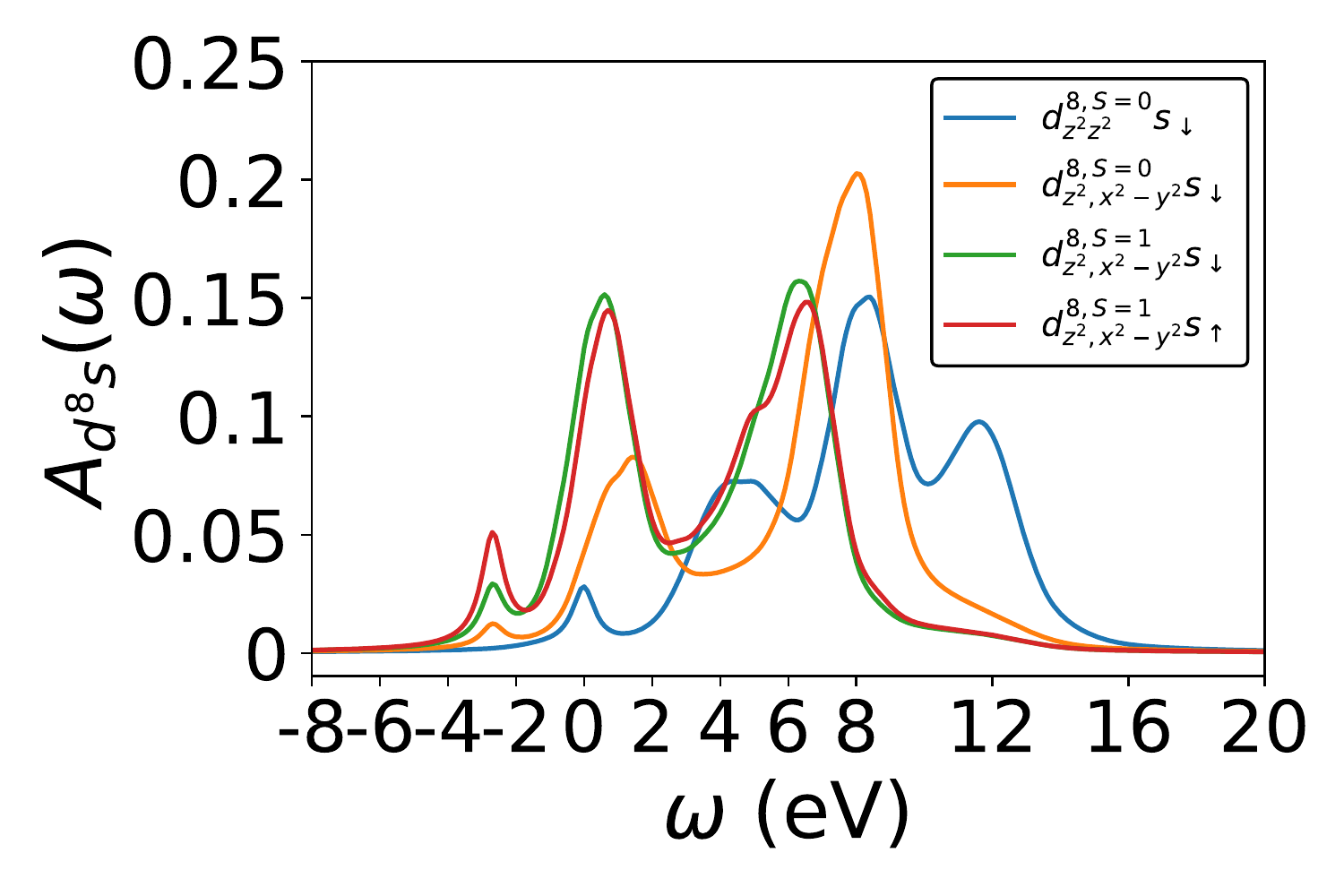,width=.65\columnwidth, trim={0 1.5cm 0 0.3cm},clip=true}
   \psfig{figure=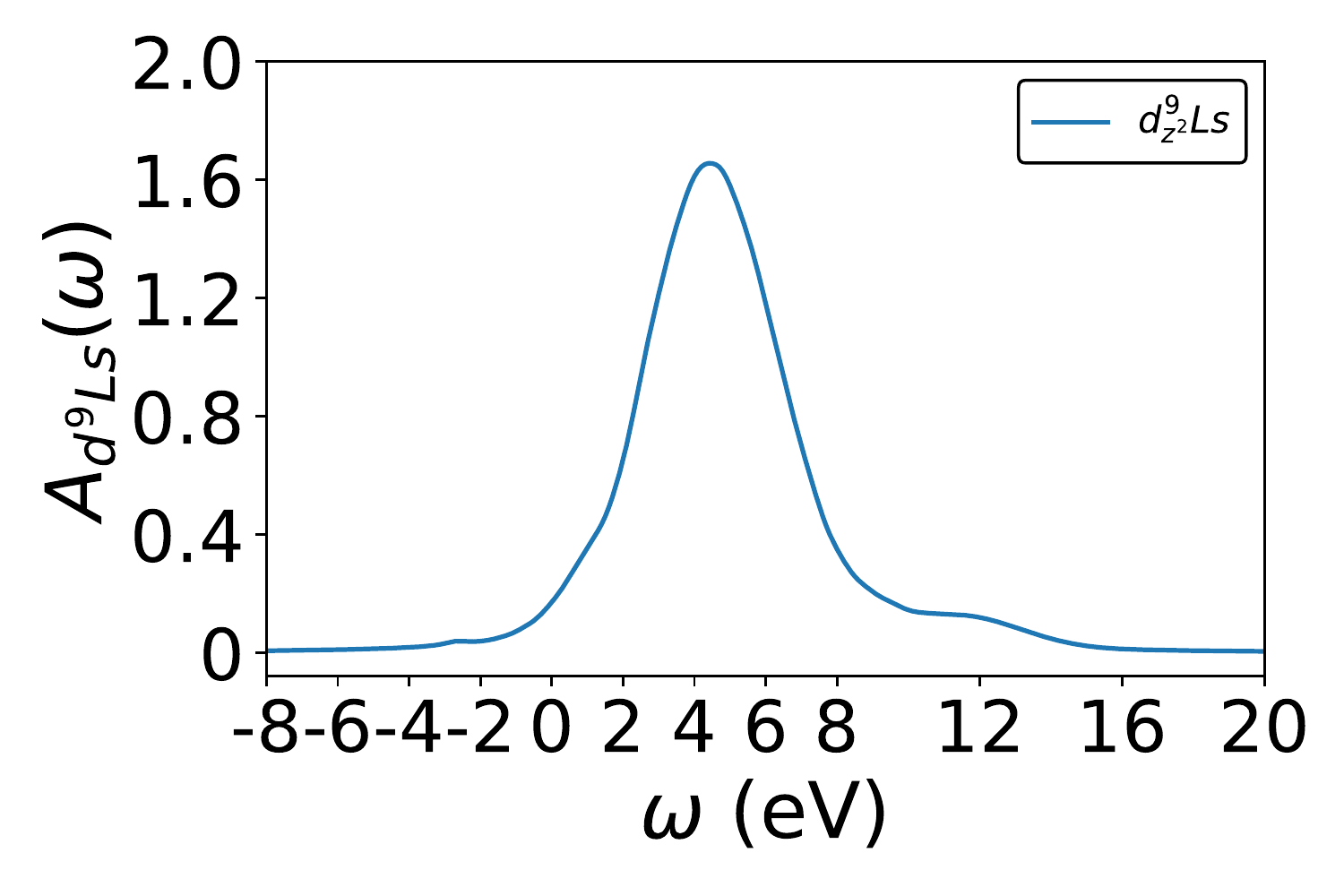,width=.65\columnwidth, trim={0 1.5cm 0 0.3cm},clip=true} 
   \psfig{figure=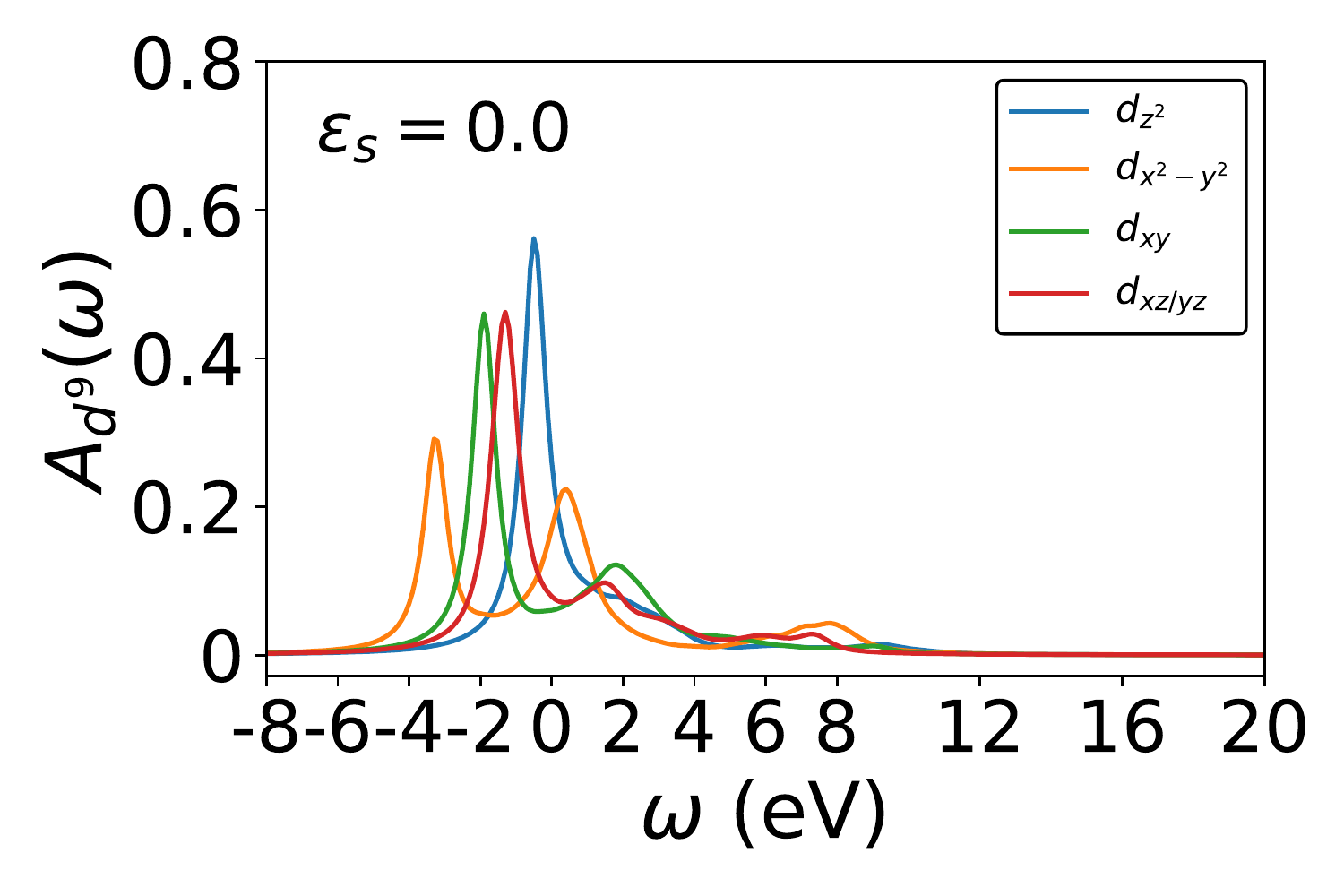,width=.65\columnwidth, trim={0 1.5cm 0 0.3cm},clip=true}
   \psfig{figure=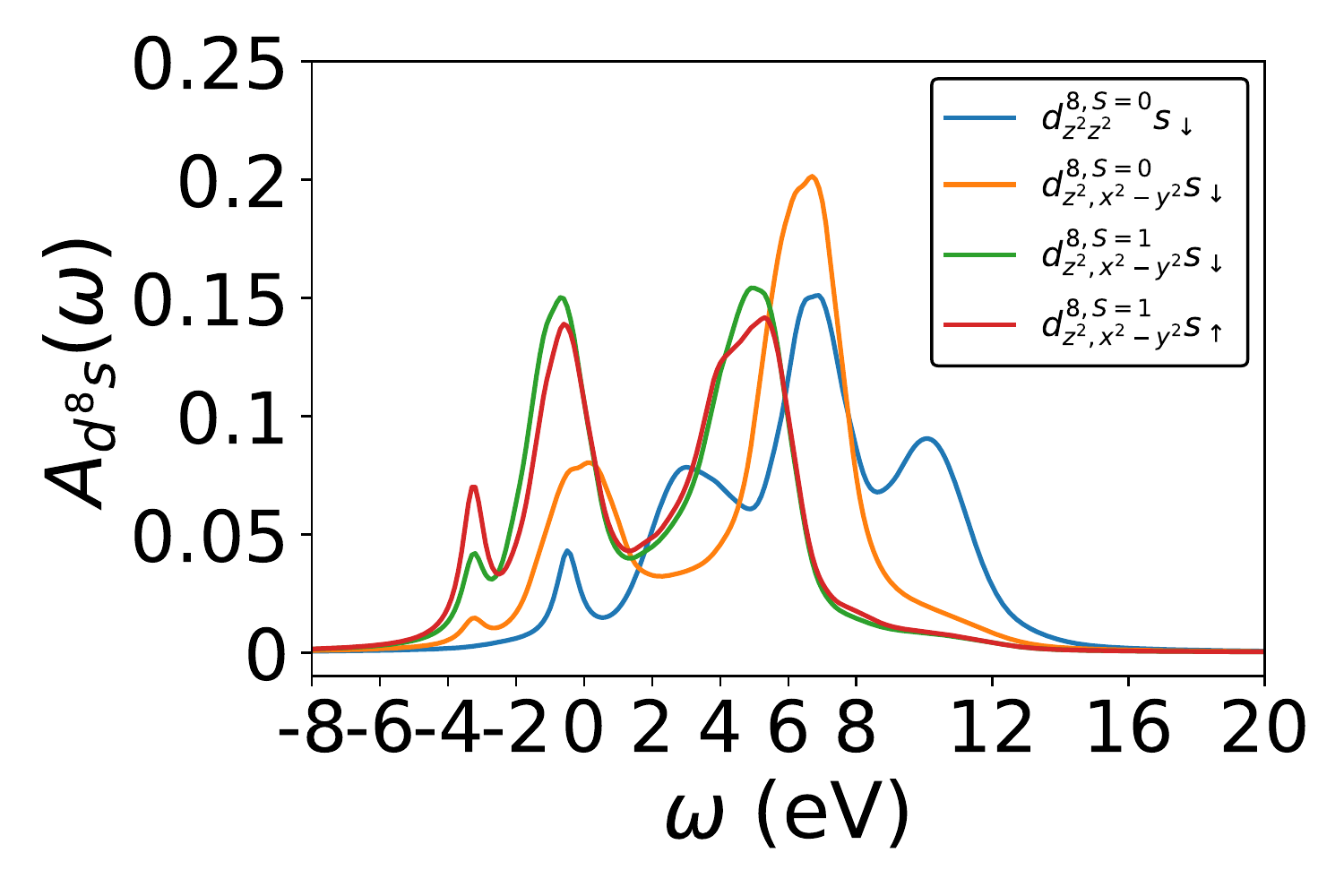,width=.65\columnwidth, trim={0 1.5cm 0 0.3cm},clip=true}
   \psfig{figure=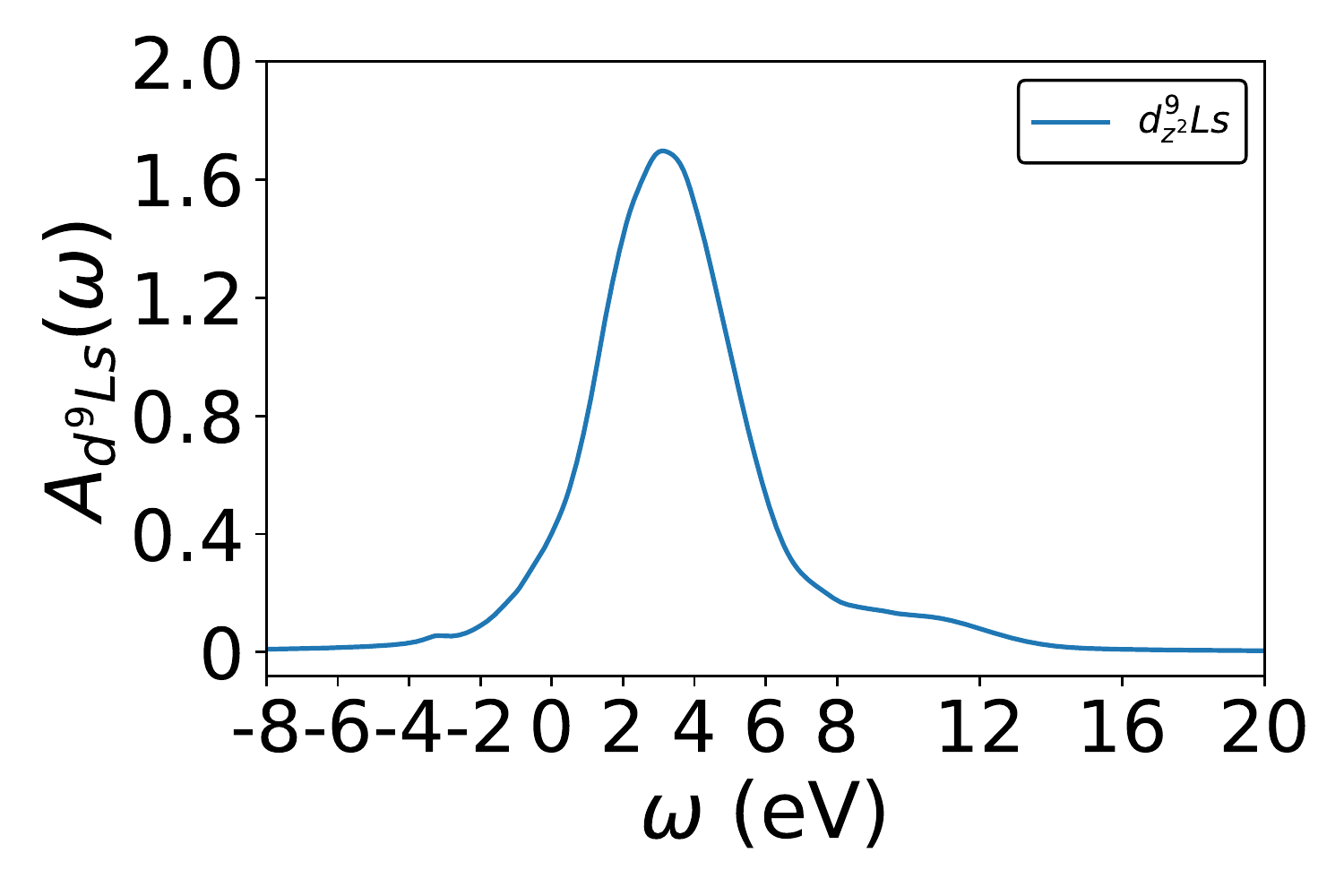,width=.65\columnwidth, trim={0 1.5cm 0 0.3cm},clip=true}
   \psfig{figure=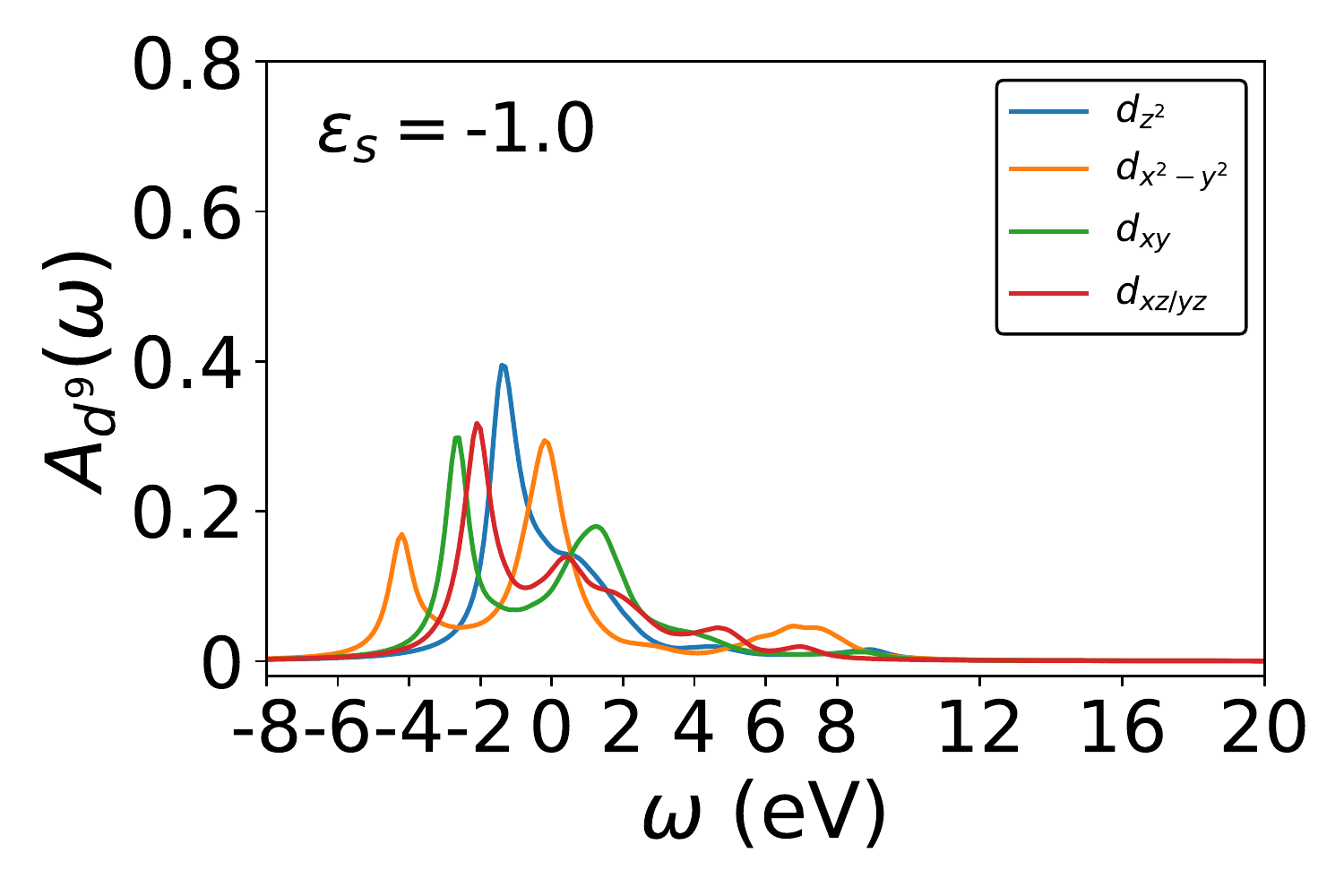,width=.65\columnwidth, clip=true}
   \psfig{figure=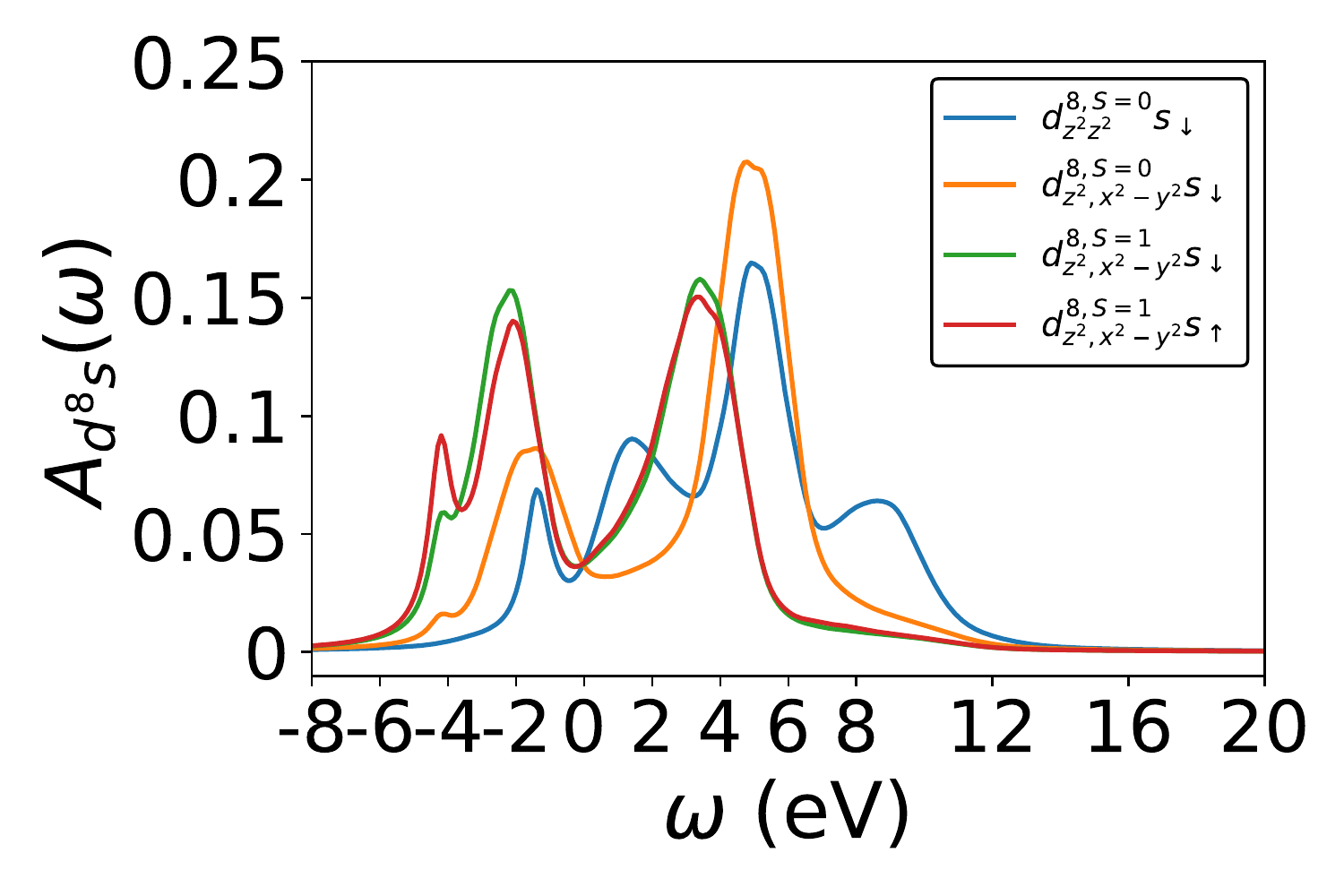,width=.65\columnwidth, clip=true}
   \psfig{figure=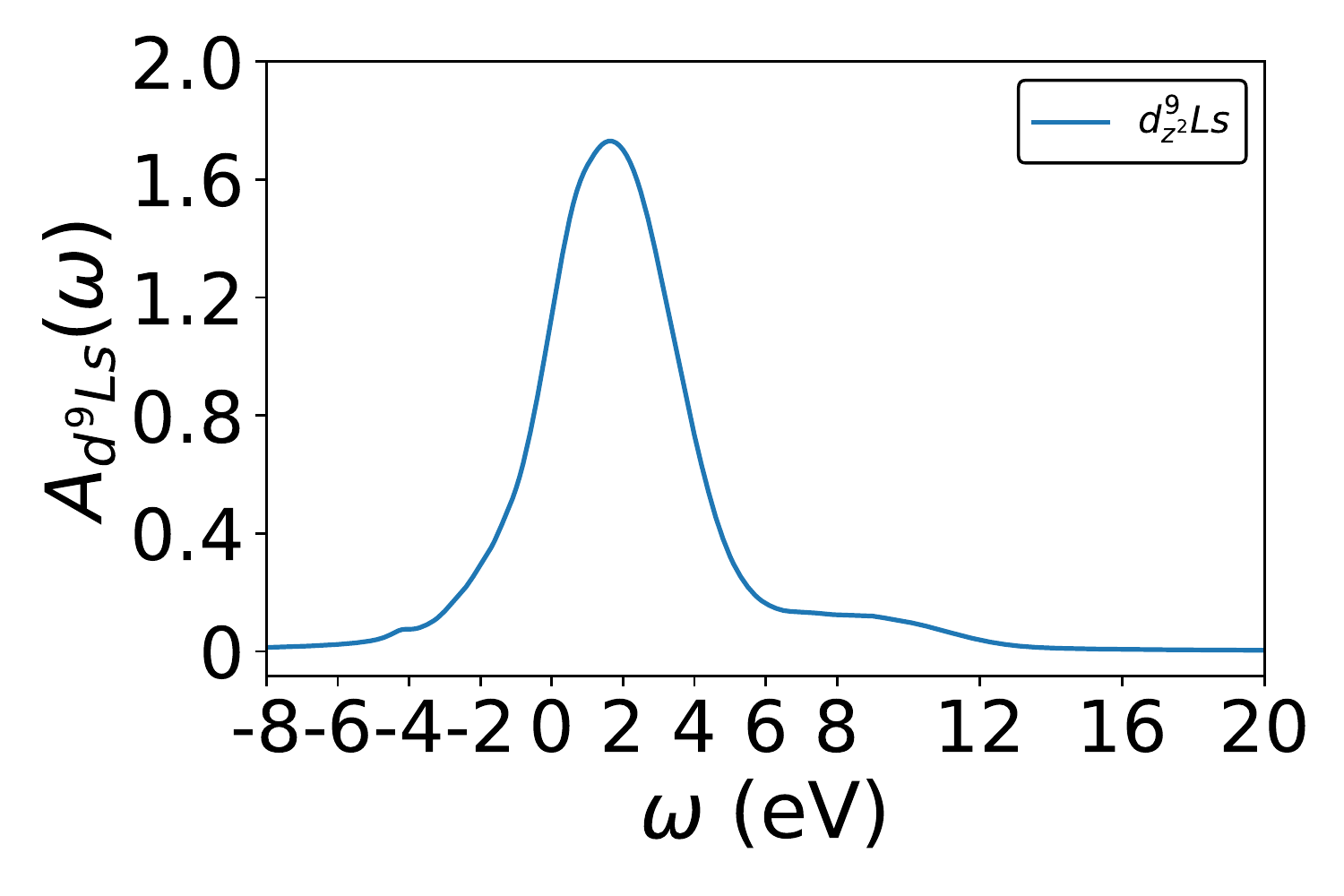,width=.65\columnwidth, clip=true} 
   \psfig{figure=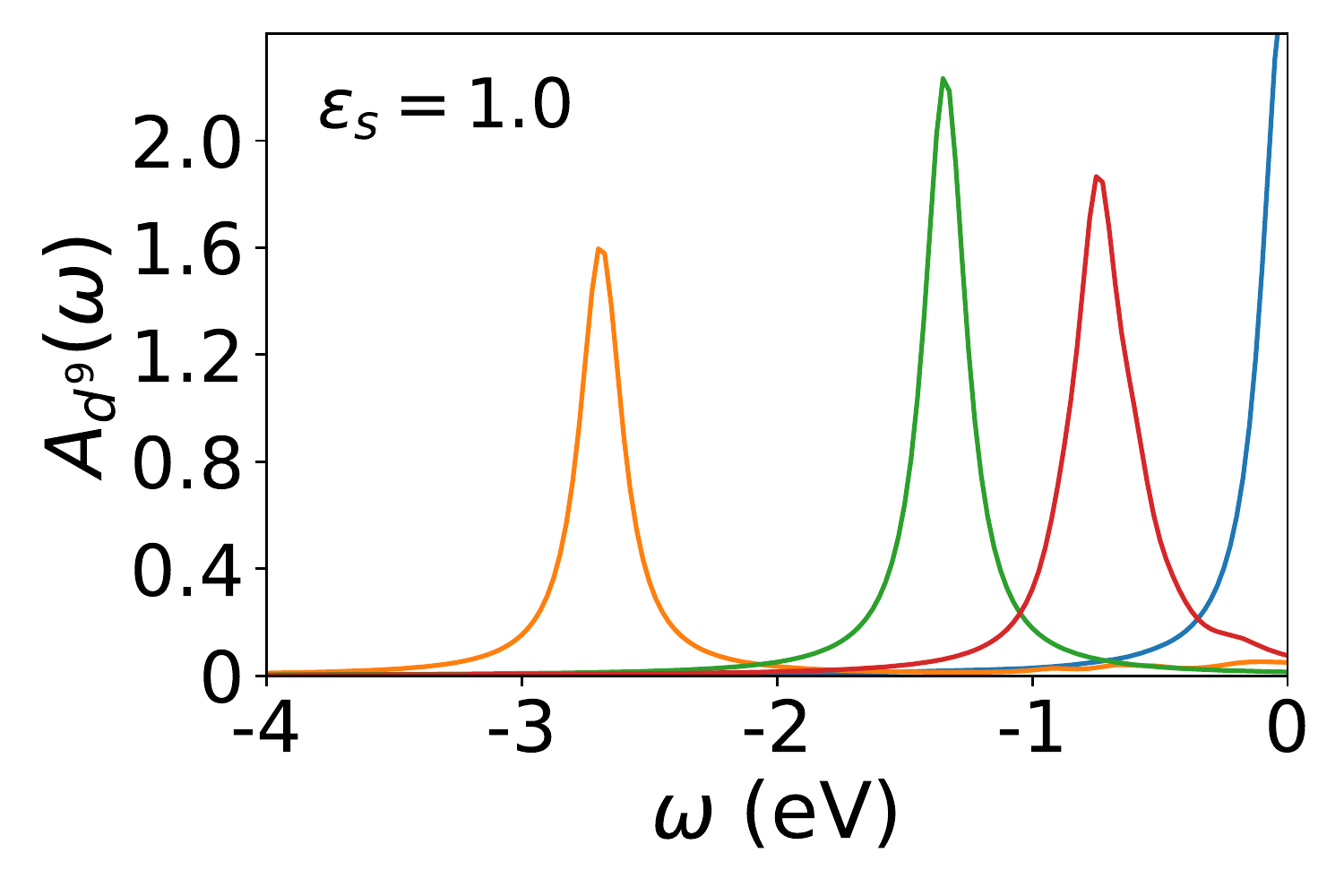,width=.65\columnwidth, clip=true}
   \psfig{figure=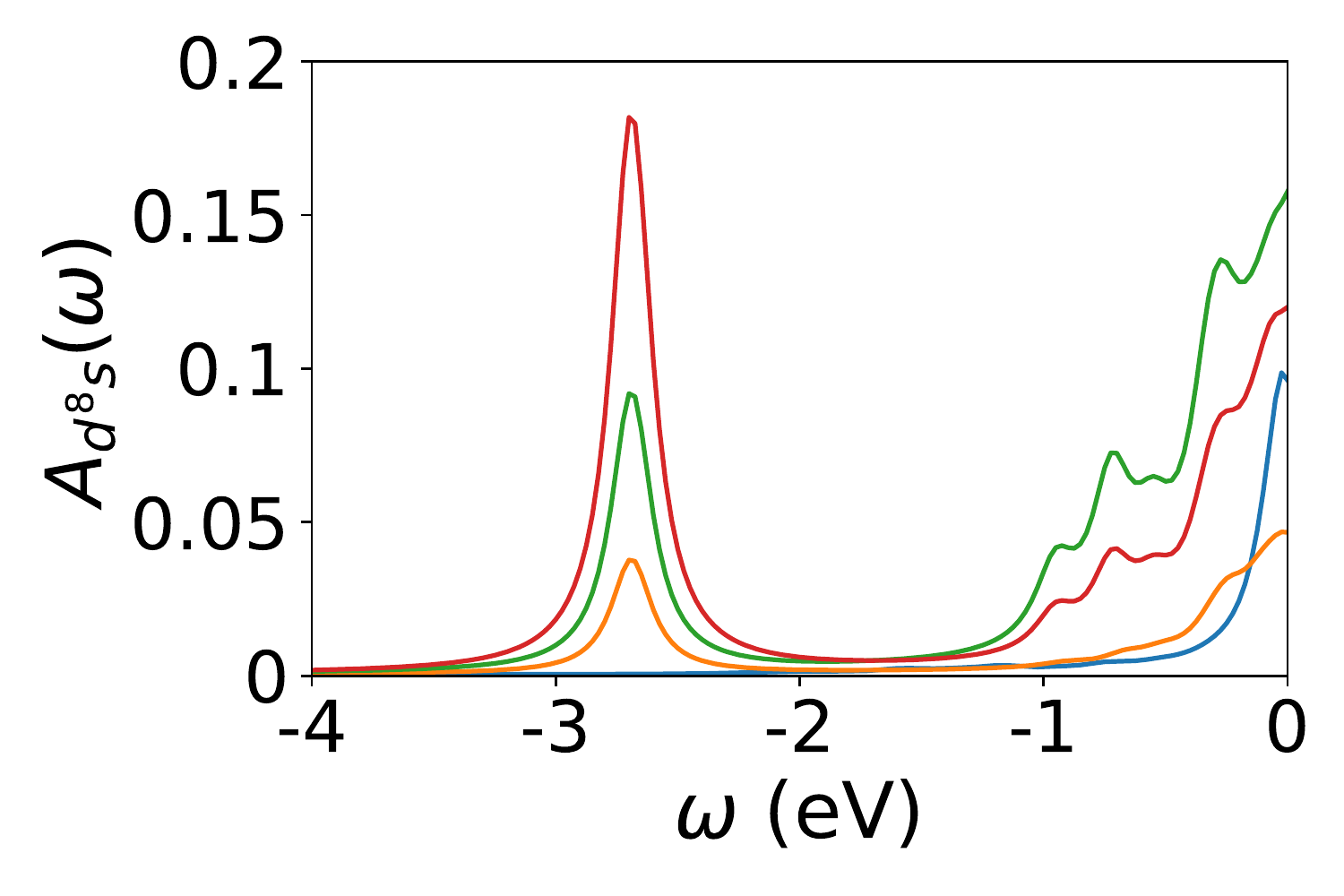,width=.65\columnwidth, clip=true}
   \psfig{figure=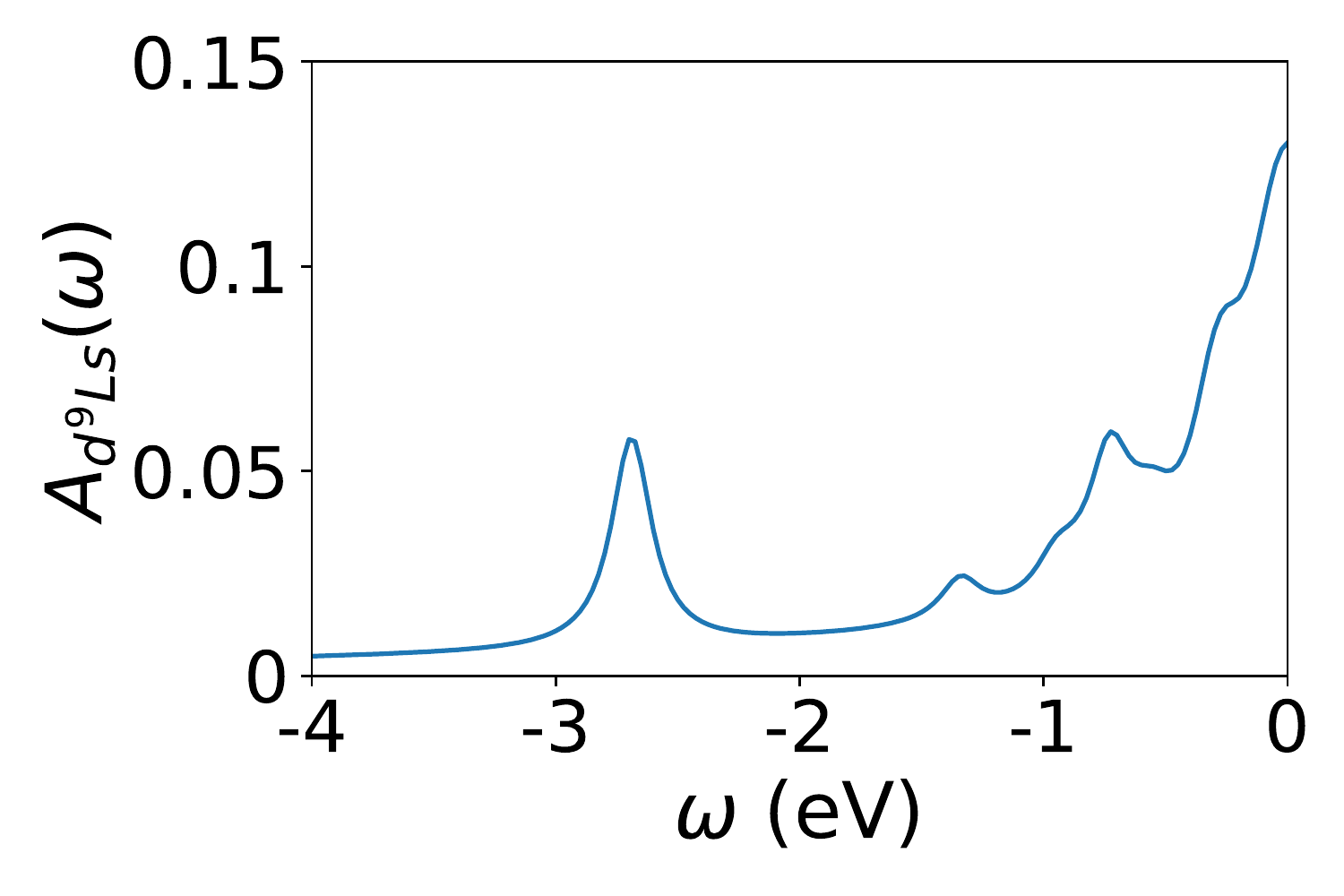,width=.65\columnwidth, clip=true} 
\caption{CS$_{N}$ spectral weights projected onto $d^9$ orbitals (left column), a variety of $d^8s$ states as indicated in the legend (middle column), and $d^9_{z^2}Ls$ states (right column). All energies are in eV. (top row) $\epsilon_s=2, \epsilon_{z^2}=1.2, \epsilon_{xy}=0.3, \epsilon_{xz/yz}=0.7$; (2nd row) $\epsilon_s=1,  \epsilon_{z^2}=0.85, \epsilon_{xy}=0.15, \epsilon_{xz/yz}=0.5$; (3rd row) $\epsilon_s=0,  \epsilon_{z^2}=0.55, \epsilon_{xy}=0.0, \epsilon_{xz/yz}=0.2$; and (4th row) $\epsilon_s=-1, \epsilon_{z^2}= \epsilon_{xy}=\epsilon_{xz/yz}=0.0$. The other parameters are  $t_{pd\sigma}=1.5, t_{pd\pi}=0.65, t_{ds}=1.13, t_{pp\sigma}=0.9, t_{pp\pi}=0.2, t_{ss}=0.23, t_{ss\perp}=0.44, \epsilon_p=3.0, \epsilon_{x^2-y^2}=0, A=6, B=0.15, C=0.58$. A broadening energy $\eta = 0.4$ has been used throughout. (bottom row) Zoomed in spectra near the ground-state energy and with a smaller broadening, for $\epsilon_s=1$. $r=0$ denotes that the excited electron in the Zs band is located right above or below Ni impurity.}
\label{fig4n}
\end{figure*}

(ii) {\bf The CS$_{N+1}$ manifold}: In the absence of Ni-Zs hybridization, this would include only the $d^{10}$ states. Hopping of an electron into the Zs band links it to the $d^9_{z^2}s$ continuum, centered at $\epsilon_s$ and with bandwidth $8t_{ss}$. This continuum is then linked through Ni-O hybridization to the $d^{10}L_{z^2}s$ continuum, centered at $U/2+\Delta+\epsilon_s$ and with a bandwidth given by the convolution of the O and Nd bands. Note that since $L$ state can only be of $3z^2-r^2$ symmetry here, this linear combination of O-2p orbitals has a different energy than the one of $x^2-y^2$ symmetry because of the influence of $t_{pp}$. Again, we ignore higher energy states with two or more electrons in the Zs band. 

(iii) {\bf The CS$_{N-1}$ manifold}: This is the most complex manifold. In addition to the $d^8$ multiplet, our previous calculation~\cite{Mi2020} included only the $d^9L$ and the $d^{10}L^2$ states. The former continuum is centered at $\epsilon_p$ and has the bandwidth of the O band, while the latter is centered at $\Delta+\epsilon_p$ and its bandwidth is doubled because there are two holes in the O band.

By emptying a $d_{z^2}$ orbital, Ni-Zs hopping  links $d^9L$ states to $d^8Ls$ states. For example, the important configuration $d^9_{x^2-y^2}L_{x^2-y^2}$ hybridizes with $d^8_{x^2-y^2,z^2}L_{x^2-y^2}s$ states forming a continuum centered at $U/2+\epsilon_p+\epsilon_s$ and whose bandwidth is given by the convolution of the O and Nd bands. Similarly, $d^{10}L^2$ states are linked to $d^9_{z^2}L^2s$ states, centered at $2\epsilon_p+\epsilon_s$ and with a bandwidth double that of the O band. Higher energy configurations are ignored.

\subsection{Undoped NdNiO$_2$: CS$_N$ spectra and GS}

To investigate the impact of the hybridization between Ni-$3d_{z^2}$ and Zs on the undoped ground state, we perform the Ni impurity calculation choosing the hopping integral between Ni-$d_{z^2}$ and Zs to be $\sim 1.13$ eV as estimated by DFT~\cite{dft25,Kat}.
To  account for the significant dispersion of $s$ band crossing the Fermi level, we follow DFT and set the $s$-$s$ hoppings to have the intra-plane value $t_{ss}=0.25$ eV and inter-plane value $t_{ss\perp}=0.44$ eV.

This still leaves as free parameters the energy $\epsilon_s$ of an electron in the Zs band, as well as the crystal field splittings of the  other four Ni-$3d$ orbitals (we set $\epsilon_d(d^9_{x^2-y^2})=0$ as the reference). These parameters should be adjusted so that the  undoped CS$_N$ spectra agree with the XAS/RIXS experiment~\cite{oneband}. The latter sets the values of the 3 splittings between peaks of different symmetries (see Fig.~\ref{fig2}) but this is  not enough to uniquely identify the values of all the free parameters. In the following we analyze a few possible values $\epsilon_s=2, 1, 0, -1$ eV while tuning (for each $\epsilon_s$) the crystal fields until we obtain the correct splittings. Their corresponding values are indicated in the caption of Fig. \ref{fig4n}.

The panels in the left column of Fig.~\ref{fig4n} show the corresponding CS$_N$ spectral weight projected onto various $d^9$ states, while those in the middle column are projected onto various $d^8s$ states, with various total spins $S,S_z$ as indicated in the legend. The right column shows projections onto $d^9_{z^2}Ls$ states. 

The left panels clearly show that for all these $\epsilon_s$ values, the ground-state has $x^2-y^2$ symmetry with considerable $d^9_{x^2-y^2}$ weight. However, as shown in the middle column panels, the ground-state also has considerable weight in the $d^8_{x^2-y^2,z^2}s$ configuration that $d^9_{x^2-y^2}$ hybridizes with via the $d^9_{z^2}$-Zs hopping. This is a key result which we will return to after we  analyze more carefully  these spectra.


%

\begin{figure}[t]
\psfig{figure=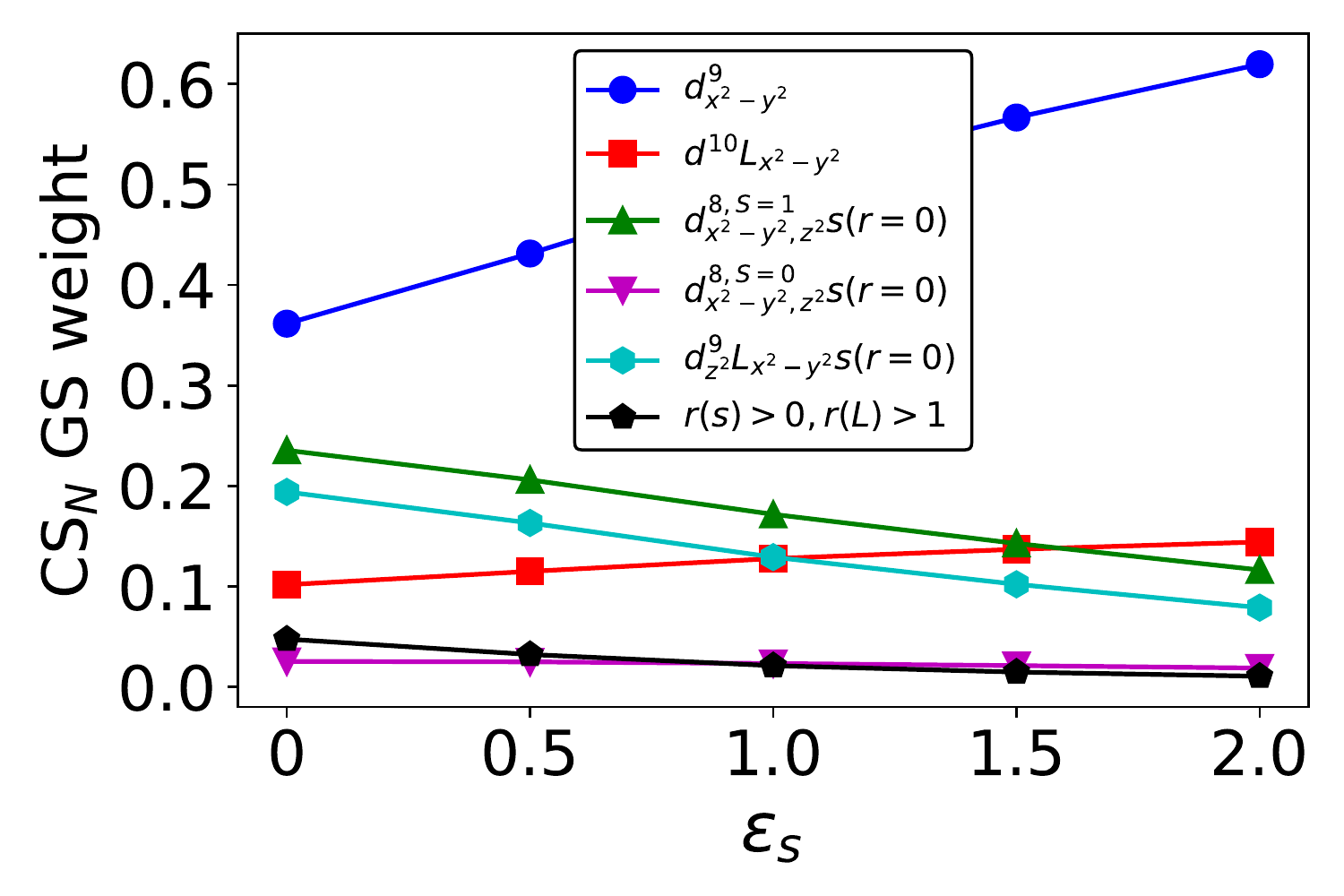, height=5.5cm,width=.95\columnwidth}
\caption{CS$_{N}$ ground state weights of dominant components. $r=0$ denotes that the excited electron in the Zs band is located right above or below Ni impurity, while $r(s)>0$ shows weights of configurations with the $s$ electron moved away from the Ni impurity. The crystal fields for each $\epsilon_s$ are adjusted for reasonable $d^9$ splitting similar to Fig.~\ref{fig4n} to be consistent with XAS/RIXS experiments.}
\label{d9GS}
\end{figure}

For the larger value $\epsilon_s=2$eV in the top-left panel, the low-energy peaks are quite similar to those in Fig.~\ref{fig2}(b), which corresponds to  $\epsilon_s \rightarrow \infty$.
  However, the spectral weights also have some higher energy features, especially visible in the $x^2-y^2$ channel which has a second peak around 2eV and a broader feature around 8-10eV. As $\epsilon_s$ is decreased (rows 2-4), most features move towards lower energies and the spectra exhibit more structure at intermediate and higher energies. This is to be expected. Fig.~\ref{fig3} shows that even for $\epsilon_s=2$eV, the $d^8s$ and $d^9Ls$ continua are closer to the $d^9$ states than the $d^{10}L$ continuum, so the former must  contribute substantially to the ground-state and push it to lower energies as $\epsilon_s$ decreases. In turn, hybridization makes these continua visible  at intermediate energies, with a weight that increases with decreasing $\epsilon_s$, in agreement with the results.

  The intermediate-energy features in the ${x^2-y^2}$ channel are therefore due to the Ni-Zs hybridization involving the $d^8s$ and $d^9Ls$ continua.
This is confirmed by the results shown in the central column. The $d^9_{x^2-y^2}$ state
hybridizes with  $d^8_{z^2, x^2-y^2}s$, and indeed, we see peaks or shoulders in these spectral weights at the GS energy of the  ${x^2-y^2}$ channel). By contrast, there is no feature at this energy for the $d^2_{z^2,z^2}s$ spectral weight, consistent with the fact that it cannot hybridize with $d^9_{x^2-y^2}$ (it does hybridize with $d^9_{z^2}$, as evidenced by appearance of a low-energy peak tracking the lowest $d^9_{z^2}$ peak).  The right column of Fig.~\ref{fig4n} shows that the projection onto the $d^9_{z^2}L_{x^2-y^2}s$ configuration also has a peak at the GS energy, confirming its mixing with the $d^9_{x^2-y^2}$ state (other $d^9Ls$ configurations, not shown, do not have this peak).

In contrast, the location of the $d^{10}L$ states is not affected by the change of $\epsilon_s$. These states are most visible in the $x^2-y^2$  channel, with which they have the strongest hybridization. For our parameters, this continuum is located roughly between 5-9eV, and indeed we can see a broad peak in the  $x^2-y^2$ spectral weight at these energies in all the panels.

\begin{figure*}
\psfig{figure=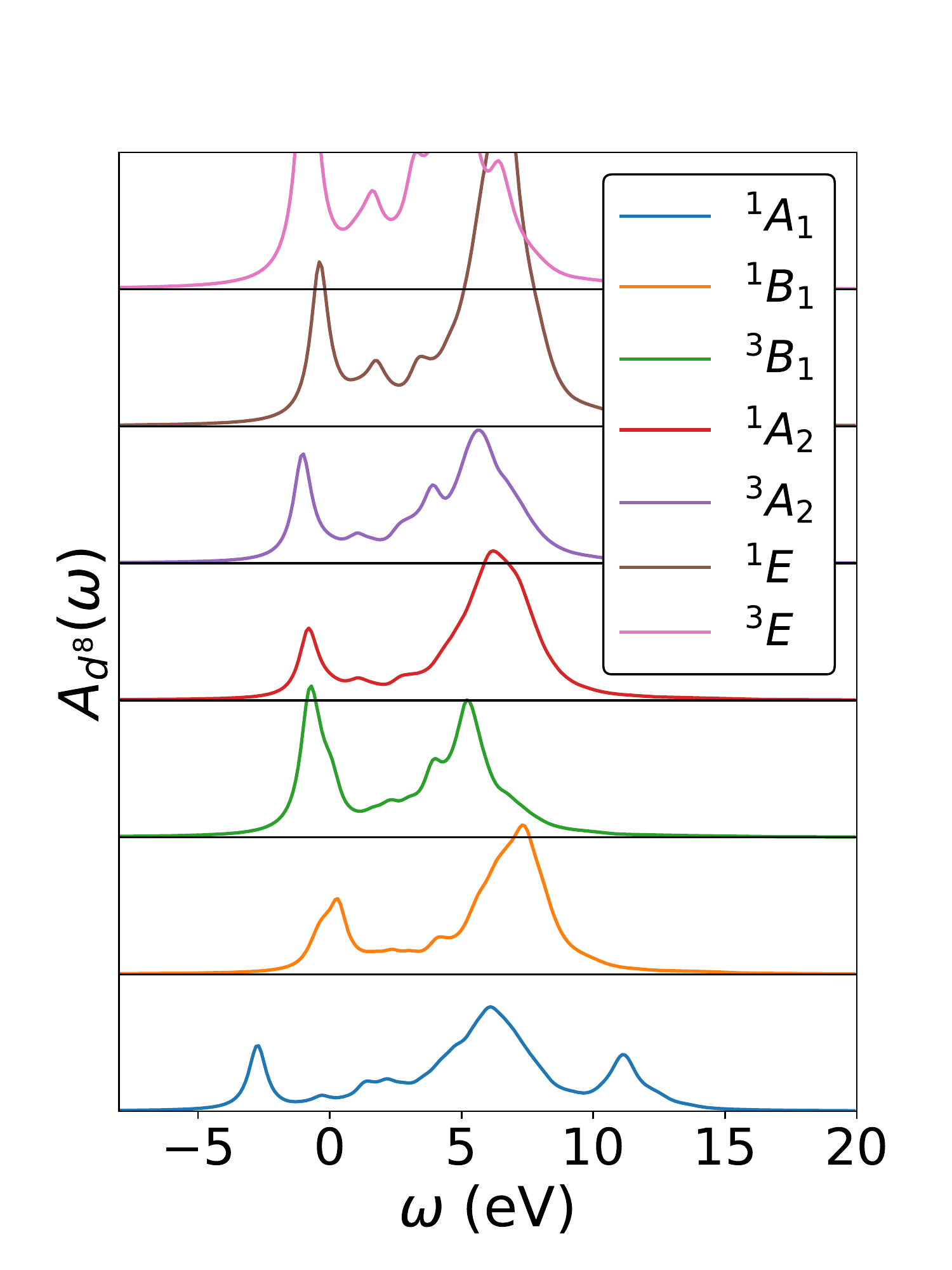,height=6.5cm,width=.65\columnwidth,trim={0 0.5cm 0 2.2cm},clip}
\psfig{figure=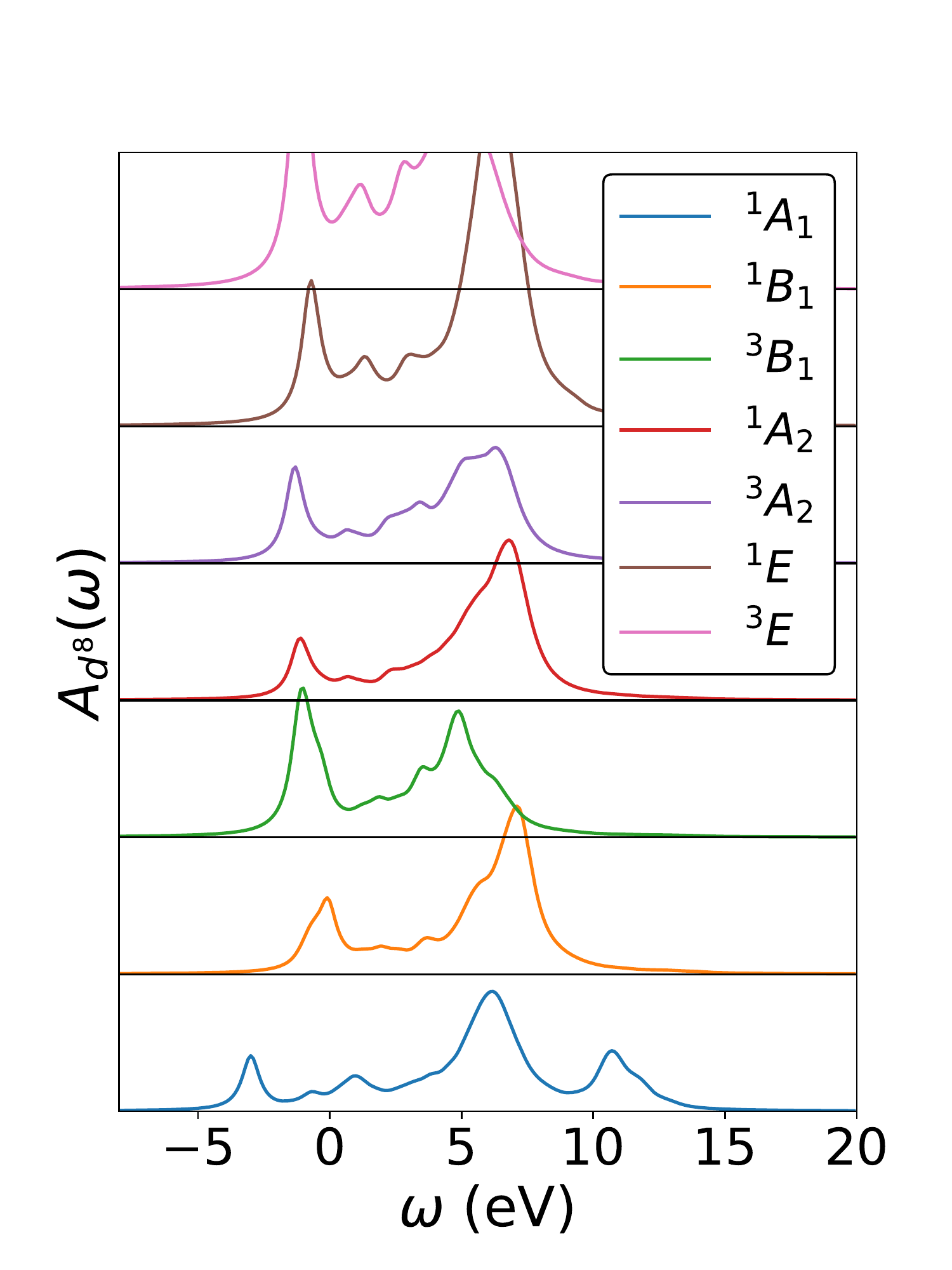,height=6.5cm,width=.65\columnwidth,trim={0 0.5cm 0 2.2cm},clip}
\psfig{figure=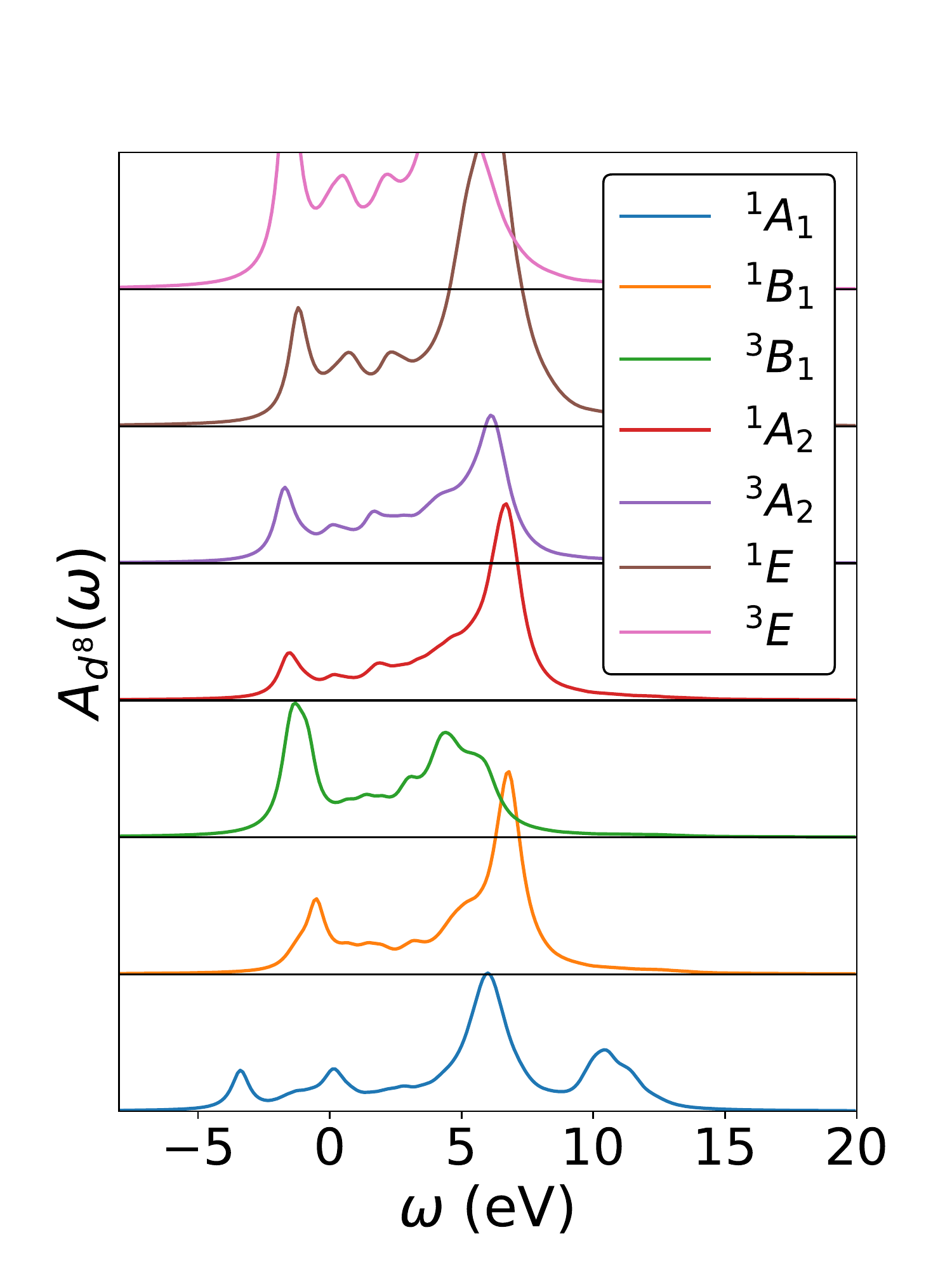,height=6.5cm,width=.65\columnwidth,trim={0 0.5cm 0 2.2cm},clip}
\psfig{figure=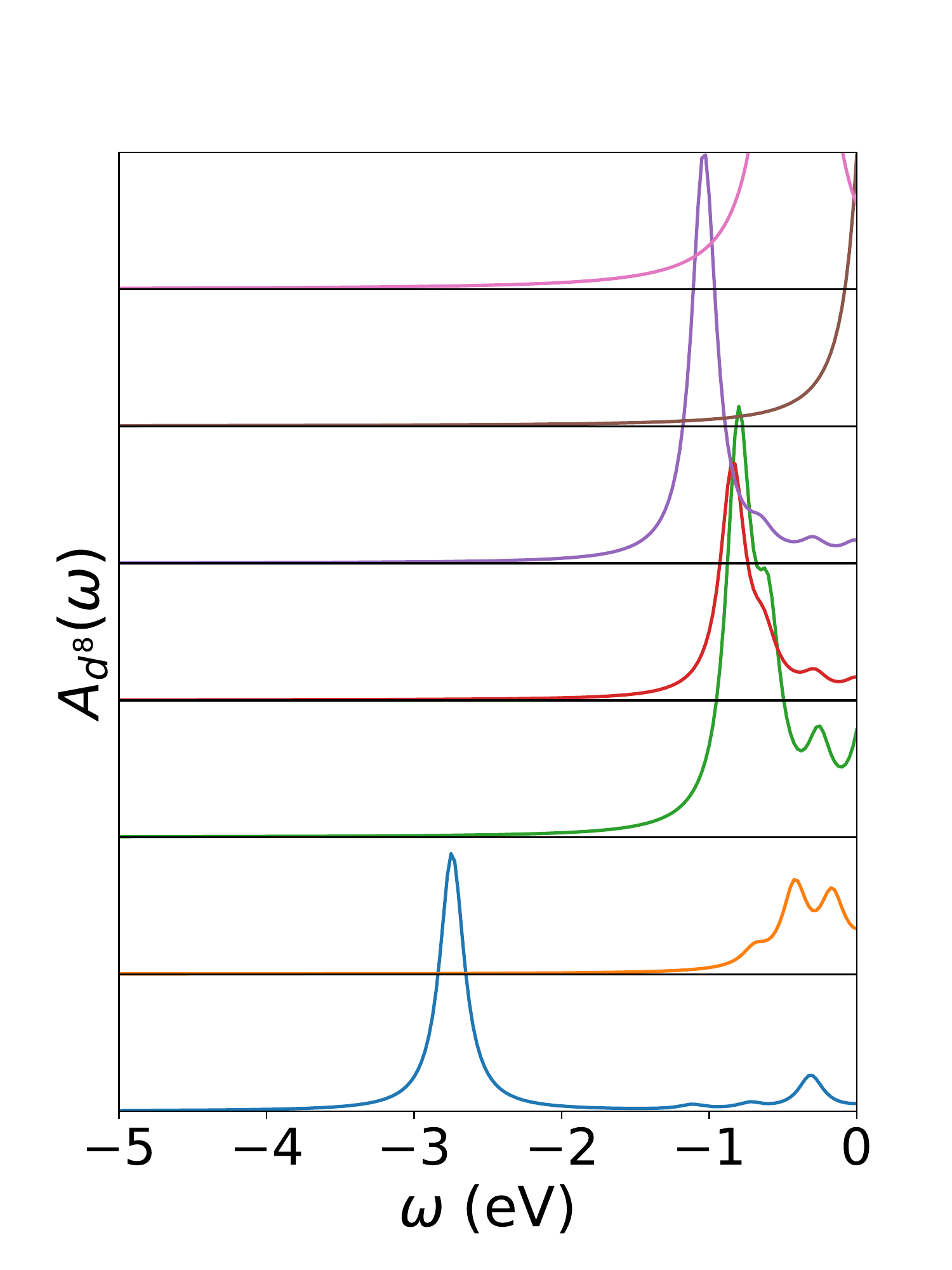,height=6.5cm,width=.65\columnwidth,trim={0 0 0 2.2cm},clip}
\psfig{figure=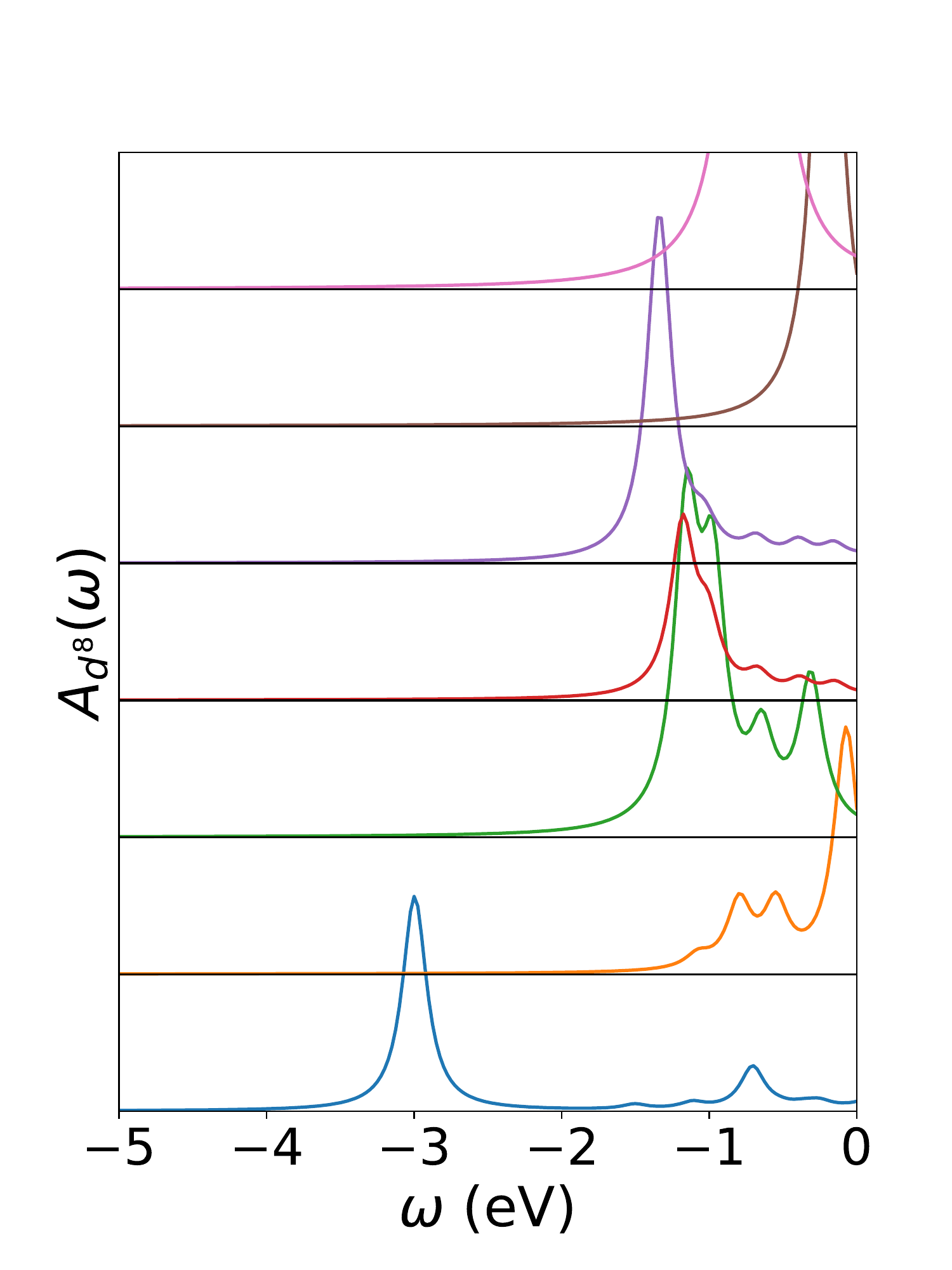,height=6.5cm,width=.65\columnwidth,trim={0 0 0 2.2cm},clip}
\psfig{figure=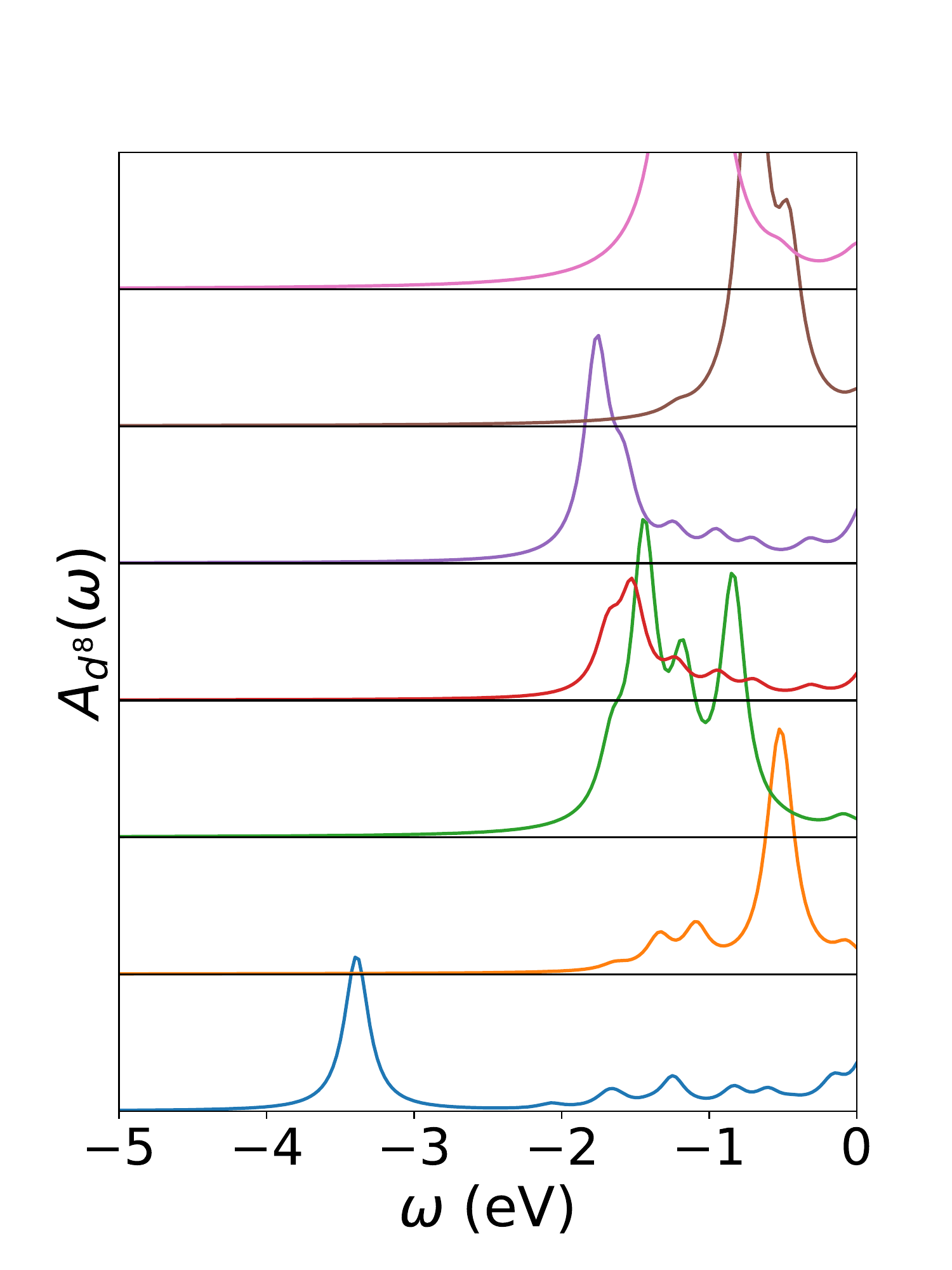,height=6.5cm,width=.65\columnwidth,trim={0 0 0 2.2cm},clip}
\caption{(Top row) CS$_{N-1}$ spectral weights projected on various $d^8$ symmetry channels. The three panels correspond to $\epsilon_s=2, 1, 0$ eV (left, middle, right, respectively). All other parameters are as used in Figure \ref{fig4n}. In all cases, the GS is in the $^1A_1$ symmetry channel. (Bottom row) Zoomed in spectra near the GS, with smaller broadening, for the same parameters as in the top row.}
\label{fig5n}
\end{figure*}

These observations are very important because they point to a crucial difference brought about by the finite Ni-Zs hybridization.  In its absence, the only way to find a partially empty $d_{z^2}$ orbital in the parent compound is through the $d^9_{z^2} \leftrightarrow d^{10}L$ hybridization in the $z^2$ channel. Experimental detection of  empty $d_{z^2}$ states would therefore be interpreted as evidence of the relevance of the $z^2$ channel. In the presence of Nd-Ni hybridization, however, our analysis reveals that experimental detection of a partially empty $d_{z^2}$ orbital is also entirely possible and expected in  the $x^2-y^2$ symmetry channel. This complicates the interpretation of experimental measurements.

The bottom row in Fig. \ref{fig4n} is a zoom  near the GS energy, of the corresponding plots from the second row, for $\epsilon_s=1$. These have a smaller broadening $\eta$ and confirm that the GS is well separated from the next higher energy states, {\it i.e.} it is indeed a discrete peak as opposed to a resonance at the bottom of a broad continuum (which the results with the larger $\eta$ might incorrectly suggest).

Figure~\ref{d9GS} further illustrates the evolution of the undoped ground state's composition as a function of $\epsilon_s$. Only configurations with substantial weights are shown here, and their weights add up to well over $90\%$. The remaining weight is distributed amongst the roughly 37 million configurations included in the calculation and not shown explicitly in this plot.

Clearly, $d^9_{x^2-y^2}$ has the dominant character in the undoped GS. 
Nonetheless, decreasing $\epsilon_s$ promotes the electron-hole pair excitation from the $d_{z^2}$ into the $s$ band, explaining the increasing contributions from the $d^8s$ and $d^9Ls$ states with decreasing $\epsilon_s$. Note that $(r=0)$ means that those weights are projected on configurations where the electron in the  Zs band is restricted to be right above or below the Ni impurity, whereas $r(s)>0$ is for configurations where the electron has moved away from the Ni impurity. Similarly, the shown configurations with an L ligand hole assume that it is on the O neighboring the Ni impurity; contributions with the ligand hole at $r(L)>1$ are very small, as shown by the black pentagons. Furthermore, we see that the two Ni-$d^8$ holes of ${x^2-y^2}$ and $z^2$ symmetries predominantly form a triplet state ($S=1$) instead of a singlet ($S=0$), in agreement with Hund's rule.
Note that as $\epsilon_s$ turns to somewhat unphysical negative values such as $\epsilon_s=-1$eV (not shown here), the GS weights of all these states become much more evenly distributed, for example the L hole can be located far away from the Ni impurity.

Fig.~\ref{d9GS} clearly shows that the total admixture of configurations involving a $d_{z^2}$ hole contribution to the $N$ particle ground state is quite large. The $d^8$ configurations involve mostly $d_{z^2}$ and $d_{x^2-y^2}$ orbitals so that both in-plane and out-of-plane polarization will be active in XAS. However, these contribution will be shifted in energy relative to the rather sharp $x^2-y^2$ dominated peak because of the electron-hole excitation left behind once one of the $d$ holes has been filled by a core electron. In principle, we could calculate spectroscopies like XAS and RIXS but this requires the inclusion of the important interaction with the core hole, and is a study in progress.

\subsection{Hole-doped NdNiO$_2$: CS$_{N-1}$ spectra and GS}

Next we diagonalize the Hamiltonian in the CS$_{N-1}$ manifold to see whether inclusion of the Ni-Zs hybridization affects its GS symmetry. The results for three typical $\epsilon_s$ values and their corresponding crystal fields are shown in Fig.~\ref{fig5n}. For simplicity, here we plot only the spectral weight projected onto various $d^8$ symmetry channels. It is clear that in all cases, the doped GS retains the $^1A_1$ symmetry.

A detailed analysis of the projections onto other configurations (not shown) confirms the charge transfer like $d^8_{^1A_1} \leftrightarrow d^9_{x^2-y^2}L  \leftrightarrow d^{10}L^2$ hybridizations similar to those dominant in a cuprate layer, confirming the conclusion of Ref. \onlinecite{Mi2020}. This is also consistent with expectations based on Fig. \ref{fig3}, according to which the lowest states in this manifold are of $d^9L$ origin. On the other hand, the $d^9Ls$ and $d^8Ls$ states with partial Zs occupation are at higher energies, and also only indirectly linked to  $d^8_{^1A1}$ through $d^9_{x^2-y^2}L \leftrightarrow d^8_{z^2, x^2-y^2}Ls$ and $d^{10}L^2\leftrightarrow d^9_{z^2} L^2 s$. This explains why the Ni-Zs hybridization has less effect on the GS of this doped manifold, as opposed to that of the undoped GS. 

\begin{figure}[t!]
\psfig{figure=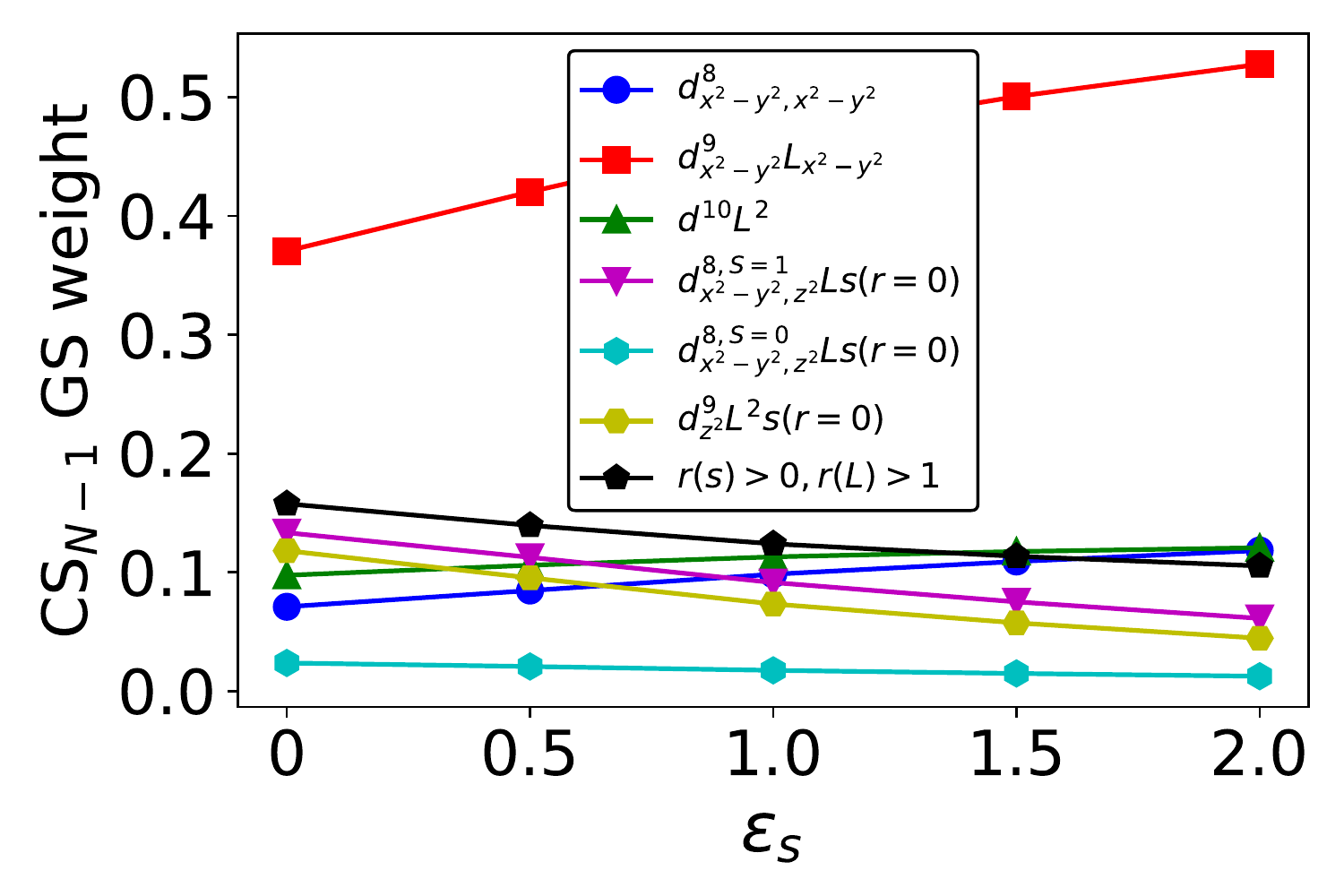, height=5.5cm,width=.95\columnwidth}
\caption{CS$_{N-1}$ ground state weights of dominant components in the $^1A_1$ channel. Notations and parameters are similar to Fig.~\ref{d9GS}.}
\label{d8GS}
\end{figure}

Similar to what was shown for the CS$_N$ GS in Fig. \ref{d9GS}, Fig.~\ref{d8GS} demonstrates the doped ground state composition as a function of $\epsilon_s$.  The dominant state is the  $d^9_{x^2-y^2}L$ singlet state, without an electronic excitation into the Zs band, regardless of $\epsilon_s$. Decreasing $\epsilon_s$ promotes the electron-hole pair excitation due to Ni-Zs hybridization, and the weights of these states become comparable or larger than that of the $d^{10}L^2$ and $d^8$ configurations, showing their importance for a quantitative description of the system. Nevertheless, we find that this doped state still looks qualitatively similar to the ZRS of the cuprate layer.

\section{Summary and outlook} 
In summary, we have adopted a Ni impurity model to explore the nature of the parent compound and hole doped states of (La, Nd, Pr)NiO$_2$ by including the crystal field splitting, the Ni-$3d$ multiplet structure, and the hybridization between Ni-$3d$ orbitals and the Nd-$5d$ orbitals mimicked by symmetric orbitals centered at the missing O in the Nd layer, forming a two-dimensional (2D) band. 
The extension to our previous work is the focus on the impact of these additional, more realistic ingredients on describing the infinite-layer nickelates, in particular the ``critical'' character of the doped hole singlet state found previously. 

First we considered the effect of only adding crystal field splittings of the $3d$ orbitals on both the undoped and one-hole doped ground-state, in the absence of involvement of the Zs orbitals. This is similar to the approximation made in the recent extended quantum chemistry calculation~\cite{Alavi} where all the hybridization involving Nd plane orbitals was neglected. We found that the presence of the crystal fields further stabilizes the hole-doped singlet state, so that the infinite-layer nickelates are more similar to the cuprates in terms of the nature of doped hole states. 

The experimental observation that the $d_{z^2}$ state spreads out over a large energy range motivated us to further couple the $d_{z^2}$ orbitals to a Zs dispersive band, associated with the Nd layers.
For the parent compound, we found that the Ni-Zs hybridization indeed results in the states of Ni-$3d^9_{z^2}$ character spreading out over a large energy range in the spectra, in qualitative agreement with  recent XAS and RIXS data.
We emphasize again that these results are expected, given that the Ni-Zs hybridization comes from an electron hopping between Ni-$d_{z^2}$ and Zs orbitals. As a result, $d^9_{x^2-y^2}$ indeed hybridizes with $d^8_{z^2,x^2-y^2}s$ states of all possible spins. It also hybridizes with $d^9_{z^2}Ls$ but indirectly, proceeding through an intermediate state $d^9_{x^2-y^2}\rightarrow d^{10}L \rightarrow d^9_{z^2}Ls$, where  the ligand hole $L$ has ``inherited'' the $x^2-y^2$ symmetry.

Our calculations pointed out that the shape of the Ni-$3d^9_{z^2}$ related structure is rather complicated, requiring reinterpretations of the experimental measurements. 
Specifically, there exists a crucial difference brought about by the finite Ni-Zs hybridization. In its absence, the only way to find a partially empty $d_{z^2}$ orbital in the parent compound is through the $d^9_{z^2} \leftrightarrow d^{10}L$ hybridization in the $z^2$ channel. Experimental detection of  empty $d_{z^2}$ states would therefore be interpreted as evidence of the relevance of the $z^2$ channel. In the presence of Ni-Zs hybridization, however, our analysis revealed that experimental detection of a partially empty $d_{z^2}$ orbital is also entirely possible and expected in  the $x^2-y^2$ symmetry channel. 

Furthermore, for the hole-doped system we showed that the inclusion of crystal fields and of the Ni-Zs hybridization still favors a lowest hole doped states of  singlet character, regardless of the site energy level of Zs orbital.  We conclude that  this doped GS is qualitatively like that in cuprates, although there are considerable quantitative differences due to the different charge transfer energy and the hybridization with the Zs band. 
The presence of $3d_{z^2}$ holes in the lowest energy hole state makes Ni look more like Ni$^{2+}$ but apparently low spin. 
The role and importance of these differences, and their effects on the superconducting state, need to be further studied.

\textcolor{blue}{}

\begin{acknowledgments}
M. Jiang is supported by the National Natural Science Foundation of China (NSFC) Grant No. 12174278, the startup fund from Soochow University, and Priority Academic Program Development (PAPD) of Jiangsu Higher Education Institutions. M. Berciu and G. A. Sawatzky are funded by the Stewart Blusson Quantum Matter Institute at University of British Columbia, and by the Natural Sciences and Engineering Research Council of Canada.
\end{acknowledgments}

\end{document}